\documentclass[acmsmall,screen]{acmart}
\usepackage{xspace}
\usepackage{cleveref}
\usepackage{listings}
\usepackage{tikz}
\usetikzlibrary{arrows.meta, positioning, fit, backgrounds, calc}
\newcommand{\name}{\textsc{GrowLibm}\xspace}
\newcommand{\acc}[2]{\ensuremath{\mathsf{acc}(#1, #2)}\xspace}

\newcommand{\pow}{\ensuremath{\operatorname{pow}}}
\newcommand{\atan}{\ensuremath{\operatorname{atan}}}
\newcommand{\atanh}{\ensuremath{\operatorname{atanh}}}
\newcommand{\logop}{\ensuremath{\operatorname{log1p}}}

\newcommand{\maxSpeedupAll}{$2.2\times$\xspace}
\newcommand{\initkStartAccuracy}{$56.0\%$\xspace}
\newcommand{\initkEndAccuracy}{$93.5\%$\xspace}
\newcommand{\maxLLVMSpeedup}{$5\%$\xspace}
\newcommand{\maxPROJSpeedup}{$1.6\times$\xspace}
\newcommand{\maxCoolPropSpeedup}{$2\times$\xspace}

\newcommand{\lowerAccuracyBasiliskSpeedup}{$2.2\times$\xspace}

\begin{document}
\title{Numerical Superoptimization for Library Learning}
\author{Jonas Regehr}
\author{Mitch Briles}
\author{Zachary Tatlock}
\author{Pavel Panchekha}

\begin{abstract}
Numerical software depends on fast, accurate implementations
  of mathematical primitives like $\sin$, $\exp$, and $\log$.
Modern superoptimizers can optimize floating-point kernels
  against a given set of such primitives,
  but a more fundamental question remains open:
  which new primitives are worth implementing in the first place?

We formulate this as \emph{numerical library learning}:
  given a workload of floating-point kernels,
  identify the mathematical primitives
  whose expert implementations
  would most improve speed and accuracy.
Our key insight is that numerical superoptimizers
  already have the machinery well-suited to this problem.
Their search procedures happen to enumerate candidate primitives,
  their equivalence procedures can generalize and deduplicate candidates,
  and their cost models can estimate \emph{counterfactual utility}:
  how much the workload would improve
  if a given primitive were available.

We present \name, which repurposes the Herbie superoptimizer
  as a numerical library learner.
\name mines candidate primitives
  from the superoptimizer's intermediate search results,
  ranks them by counterfactual utility,
  and prunes redundant candidates.
Across three scientific applications
  (PROJ, CoolProp, and Basilisk),
  \name identifies compact, reusable primitives
  that can be implemented effectively
  using standard numerical techniques.
When Herbie is extended with these expert implementations,
  kernel speed improves by up to \maxSpeedupAll
  at fixed accuracy,
  and maximum achievable accuracy also improves,
  in one case from \initkStartAccuracy to \initkEndAccuracy.
We also prototype an LLVM matcher
  that recognizes learned primitives in optimized IR,
  recovering 26 replacement sites across five PROJ projections
  and demonstrating measurable effects
  on both end-to-end speed and accuracy.\looseness=-1
\end{abstract}

\maketitle 
 
\section{Introduction}
\label{sec:intro}

Maps and navigation, physical simulation, engineering CAD---%
  computation in these increasingly important fields
  relies on short floating-point kernels whose
  speed and accuracy shape application behavior.
Those kernels, in turn, are built from the
  mathematical primitives available on the target platform.
These primitives include basic arithmetic operators
  as well as elementary functions provided by
  system math libraries like $\sin$, $\exp$, and $\log$.
They might also include third-party libraries like GSL,
  or even application-specific library functions written by experts.
Floating-point kernels then compose these primitives
  into higher-level mathematical operations
  while taking care to ensure both speed and accuracy.
  
In fact, the programming languages community
  has spent over a decade developing tooling to support this task.
Modern numerical superoptimizers like Herbie~\cite{herbie},
  for example, can increasingly optimize kernels against
  a given set of available primitives~\cite{chassis},
  even scaling the results to real-world applications
  like the LULESH hydrodynamics application~\cite{poseidon}.
Modern verification tools like Satire~\cite{satire}
  make it possible to statically verify these implementations,
  given error models for those primitives.
And modern precision tuners
  like FPTuner~\cite{fptuner} and POP~\cite{pop}
  can even optimize kernels to leverage mixed precision.
Designing a particular floating-point kernel
  against a particular set of primitives
  is increasingly well supported.

This only raises the question:
  which new primitives are worth building?
Some primitives, like $\sin$ or matrix multiplication,
  are so widely used that implementation justifies itself.
But how should library authors prioritize the zoo
  of mathematical functions found across fields
  as diverse as geospatial projections, Helmholtz energy formulations,
  astrodynamics, and more?
How do library authors determine
  which primitives are most common, and most important,
  to their particular domain?\looseness=-1

We call this problem \emph{numerical library learning}:
  given an application and a representative workload,
  identify mathematical primitives whose specialized implementations
  would most improve the application's speed/accuracy tradeoff.
Numerics are an especially attractive setting because
  primitives are small, pure expressions,
  while their implementations can range widely,
  covering algebraic identities, multi-word arithmetic,
  approximations of various types, digit recurrences,
  and even hardware-aware tuning.
In other words, primitives are easy to characterize
  but require expertise to implement well.
Numerical library learning, then, prioritizes that expertise.

Our key insight is that numerical superoptimizers
  already have the machinery of library learners.
Their search procedures enumerate plausible primitives,
  their equivalence procedures identify
    different presentations of the same latent function, and
  their error and cost models answer
    the question library designers actually care about:
    if this primitive existed,
    where would it be used,
    and how much would it help?

We present \name, a numerical library learner
  that uses a numerical superoptimizer ``in reverse'':
  instead of optimizing a kernel against a fixed set of primitives,
  it asks which new primitives would improve many kernels at once.
\name mines candidate primitives from the superoptimizer's own search space,
  semantically deduplicates and generalizes them,
  and ranks them by \emph{counterfactual utility}:
  how much the workload would improve if
  the primitive were available.
Intuitively,
  \name favors candidates that are frequent, small,
  yet still hard for a superoptimizer to optimize on its own.
The result is a short list at the right scope for expert attention:
  compact, reusable primitives that recur across kernels
  and unlock better implementations.

We evaluate \name on three substantial scientific applications:
  cartographic projections in PROJ~\cite{proj},
  thermophysical property computations in CoolProp~\cite{coolprop}, and
  astrodynamics kernels in Basilisk~\cite{basilisk}.
Across these applications,
  \name proposes compact domain-specific primitives
  that a numerical expert can implement effectively.
When implemented well,
  many of these primitives are reused in optimized kernels and
  expand the speed/accuracy frontier,
  with speed-ups of up to \maxSpeedupAll compared to
  similarly-optimized kernels without access to \name-suggested primitives.
The new primitives also enable
  higher accuracy than previously possible,
  with one kernel in PROJ going from
  \initkStartAccuracy to \initkEndAccuracy accurate
  thanks to its use of 2 \name-suggested primitives.
We also prototype an LLVM matcher that
  recognizes learned primitives in optimized LLVM IR,
  recovering 26 replacement sites across five PROJ projections.
Some projections are both faster (by up to \maxLLVMSpeedup)
  and more accurate;
  others trade modest slowdowns for substantially improved accuracy,
  illustrating how learned primitives
  expand the deployment-level speed/accuracy frontier.

\paragraph{Contributions.}
\begin{itemize}

\item We formulate \emph{numerical library learning},
  the workload-driven problem of deciding which new
  mathematical primitives are worth expert implementation
  for a given numerical application.

\item We present \name,
  which repurposes the Herbie superoptimizer as a library learner
  by mining candidates from its search space,
  semantically deduplicating and generalizing them,
  and ranking them by counterfactual utility.

\item We show that, across PROJ, CoolProp, and Basilisk,
  \name finds compact primitives
  that a numerical expert can implement effectively,
  that Herbie can reuse in optimized kernels,
  and that expand the reachable speed/accuracy frontier.

\item We demonstrate a path to deployment
  with a prototype LLVM matcher that recognizes learned primitives
  in optimized IR and rewrites them to deployed implementations.

\end{itemize}

\begin{figure}
\centering
\resizebox{\columnwidth}{!}{%
\begin{tikzpicture}[
  box/.style={
    draw, rounded corners=2pt, align=center,
    font=\footnotesize, inner sep=5pt,
    minimum height=11mm,
  },
  herbiebox/.style={
    draw=green!50!black, thick, rounded corners=2pt,
    inner sep=5pt,
    minimum height=11mm,
    fill=green!8,
  },
  iobox/.style={
    draw, rounded corners=6pt, align=center,
    font=\footnotesize\bfseries,
    inner sep=5pt, minimum height=11mm,
    fill=black!6,
  },
  phase/.style={
    draw, thick, rounded corners=3pt,
    inner sep=2.5mm,
    inner ysep=3.5mm,
  },
  phaselabel/.style={
    font=\bfseries\small,
    anchor=south west,
    inner sep=2pt,
    fill=white,
  },
  arr/.style={->, >={Stealth[length=5pt]}, semithick},
  edgelabel/.style={font=\scriptsize, align=center, fill=white, inner sep=1pt},
]


\node[herbiebox] (herbie1) at (0,0) {%
  \begin{tabular}{@{}l@{}}
  \multicolumn{1}{@{}c@{}}{\footnotesize\textbf{Herbie}} \\[-1pt]
  \scriptsize\textbf{Workload:} source kernels \\
  \scriptsize\textbf{Platform:} default
  \end{tabular}%
};

\node[box, right=10mm of herbie1] (cut) {Extract subs \&\\cut at operators};
\node[box, right=10mm of cut] (canon) {Canonicalize\\\& deduplicate};

\begin{scope}[on background layer]
\node[phase, fit=(herbie1)(cut)(canon)] (genbox) {};
\end{scope}

\node[phaselabel] at ($(genbox.north west) + (4pt, -2pt)$) {Generation};

\node[box] at ($(genbox.west |- herbie1) + (-18mm, 0)$) (convert) {Convert to\\FPCore};
\node[iobox] at ($(convert) + (-30mm, 0)$) (kernels) {Numerical\\kernels};


\coordinate (row2top) at ($(genbox.south) + (0, -14mm)$);

\node[herbiebox, anchor=north east] at ($(genbox.east |- row2top) + (-2.5mm, 0)$) (herbie2) {%
  \begin{tabular}{@{}l@{}}
  \multicolumn{1}{@{}c@{}}{\footnotesize\textbf{Herbie}} \\[-1pt]
  \scriptsize\textbf{Workload:} candidates \\
  \scriptsize\textbf{Platform:} grow
  \end{tabular}%
};

\node[box, left=12mm of herbie2] (score) {Rank by frequency,\\urgency, and size};

\node[box, below=10mm of $(herbie2)!0.5!(score)$] (addtop) {
  Add top candidates\\to platform\\(deduplicate by implication)
};

\begin{scope}[on background layer]
\node[phase, fit=(herbie2)(score)(addtop)] (selbox) {};
\end{scope}

\node[phaselabel] at ($(selbox.north west) + (4pt, -2pt)$) {Selection};

\node[herbiebox, left=10mm of selbox.west |- score] (herbie3) {%
  \begin{tabular}{@{}l@{}}
  \multicolumn{1}{@{}c@{}}{\footnotesize\textbf{Herbie}} \\[-1pt]
  \scriptsize\textbf{Workload:} source kernels \\
  \scriptsize\textbf{Platform:} grow
  \end{tabular}%
};

\node[box, left=10mm of herbie3] (filter2) {Filter by\\number\\of uses};

\node[iobox, left=10mm of filter2] (output) {Proposed\\primitives};


\draw[arr] (kernels) -- (convert);
\draw[arr] (convert) -- (herbie1);
\draw[arr] (herbie1) -- (cut);
\draw[arr] (cut) -- (canon);

\draw[arr] (canon.east) -- ++(8mm,0) |- (herbie2.east);

\draw[arr] (herbie2) -- (score);
\draw[arr] (score) -- (addtop);
\draw[arr] (addtop) -- (herbie2);

\draw[arr] (score) -- (herbie3);
\draw[arr] (herbie3) -- (filter2);
\draw[arr] (filter2) -- (output);

\end{tikzpicture}%
}
\caption{
  The \name pipeline.
  Given numerical kernels converted to FPCore~\cite{fpcore,fpbench},
    a standard floating-point interchange format,
    the \textbf{generation} phase
    runs Herbie to explore equivalent programs,
    extracts subexpressions from Herbie's intermediates,
    and canonicalizes and deduplicates them.
  The \textbf{selection} phase iteratively
    ranks candidates by frequency, urgency, and size;
    resolves implications between top candidates
    so that redundant variants are not both selected;
    and adds the winners to Herbie's platform
    for re-evaluation.
  A final Herbie pass confirms
    which candidates are actually used in optimized kernels,
    producing the proposed primitives.
}
\label{fig:pipeline}
\end{figure}
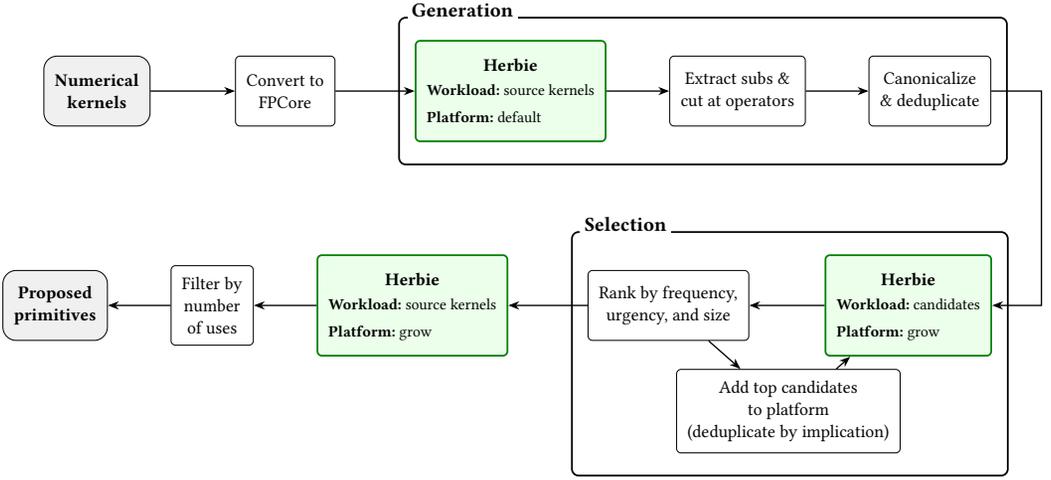

\section{Overview}
\label{sec:overview}

We illustrate numerical library learning
  with a worked example from the PROJ cartography library~\cite{proj.org},
  developed by the Open Source Geospatial Foundation~\cite{osgeo.org},
  which is used for transforming geospatial data
  between coordinate systems.
PROJ includes a large library of map projections,
  each of which defines forward and inverse projection functions.
\Cref{fig:somerc} shows an example projection function
  from the Swiss Oblique Mercator projection,
  a variation on the normal Mercator projection
  that tilts the projection cylinder so that the line of true scale
  runs west to east across Switzerland.
This (obscure) projection is the standard map projection
  for the government of Switzerland
  and thus useful for handling some Swiss government data.

\begin{figure}
\begin{lstlisting}[language=C]
static PJ_XY somerc_e_forward(PJ_LP lp, PJ *P) { /* Ellipsoidal, forward */
    PJ_XY xy = {0.0, 0.0}; double phip, lamp, phipp, lampp, sp, cp;
    struct pj_somerc *Q = static_cast<struct pj_somerc *>(P->opaque);

    sp = P->e * sin(lp.phi);
    phip = 2. * atan(exp(Q->c * (log(tan(M_FORTPI + 0.5 * lp.phi)) -
                                    Q->hlf_e * log((1. + sp) / (1. - sp))) +
                            Q->K)) - M_HALFPI;
}
\end{lstlisting}
\caption{
  The forward ellipsoidal projection function
    for the Swiss Oblique Mercator projection
    from the PROJ library~\cite{somerc.cpp}.
  Note the complex compositions of mathematical functions,
    including $2 \atan(\exp(x)) - \pi/2$,
    $\log(\tan(\pi/4 + 0.5 x))$, and
    $\log((1 + x) / (1 - x))$.
  Accurately evaluating these compositions
    is challenging in floating point
    and generally beyond the capabilities of
    existing numerical superoptimizers.
}
\label{fig:somerc}
\end{figure}

\begin{figure}
\begin{lstlisting}[language=C]
double log1pmd(double x) {
    static const double
    Lp1 = 6.666666666666735130e-01, Lp2 = 3.999999999940941908e-01,
    Lp3 = 2.857142874366239149e-01, Lp4 = 2.222219843214978396e-01,
    Lp5 = 1.818357216161805012e-01, Lp6 = 1.531383769920937332e-01,
    Lp7 = 1.479819860511658591e-01;

    double z = x * x;

    double R = fma(z, fma(z, fma(z, fma(z, fma(z, fma(z,
               Lp7, Lp6), Lp5), Lp4), Lp3), Lp2), Lp1);

    return fma(x * z, R, 2.0 * x);
}
\end{lstlisting}
\caption{
  A numerical expert's implementation of $\log((1 + x) / (1 - x))$,
    using a polynomial that is accurate for inputs $|x| < .1716$.
  This implementation borrows from
    the famous \texttt{fdlibm} math library.
  The input range is sufficient for the use case in \Cref{fig:somerc}
    because there the relevant input $x$ is
    the eccentricity of the Earth times the sine of the oblique angle,
    well within that input range.
}
\label{fig:log1pmd}
\end{figure}

This short code snippet uses several complex mathematical expressions
  like $2 \atan(\exp(x)) - \pi/2$, $\log(\tan(\pi/4 + 0.5 x))$, and $\log((1 + x) / (1 - x))$,
  which are part of the domain of coordinate system transformation.
These expressions are compiled, on typical systems,
  to calls into an underlying system math library like GLibC,
  which have typically been written by experts in math function implementation.
These system math libraries are accurate and fast,
  largely due to the efforts of those same experts
  and the broader numerical analysis community.

The problem is that, though system math libraries may be fast and accurate,
  composing accurate functions does not always result
  in accurate expressions.
In fact, all three expressions noted above---%
  $2 \atan(\exp(x)) - \pi/2$, $\log(\tan(\pi/4 + 0.5 x))$, and $\log((1 + x) / (1 - x))$---%
  are inaccurate for certain inputs:
  $2 \atan(\exp(x)) - \pi/2$ is inaccurate for $x \approx 0$,
  $\log(\tan(\pi/4 + 0.5 x))$ is inaccurate for $|x| > 1$ or $|x| \approx 0$,
  and $\log((1 + x) / (1 - x))$ is inaccurate for $x \approx 0$.
Determining whether these situations are physically realistic,
  and understanding their impact on application-level metrics
  like geospatial accuracy,
  requires substantial expertise in the domain.
If the expressions are, in fact, problematically inaccurate,
  numerical methods expertise is further required
  to fix the underlying problem.

One option for improving expression accuracy
  is any of a number of numerical superoptimizers and compilers
  developed by the programming languages community,
  including Herbie~\cite{herbie,pherbie,chassis},
  Salsa~\cite{salsa,sardana,salsa-sttt}, and Rosa~\cite{rosa,daisy}.
While these tools are convenient,
  the techniques they apply are limited
  and they are therefore less capable than a numerical analyst.
Herbie, for example,
  can propose an accurate (though slower) implementation of \texttt{log1pmd}%
\footnote{
  Herbie proposes $\logop(x) - \logop(-x)$;
    another valid implementation might be $2 \atanh(x)$.
  Neither of these implementations is
    as fast as the expert implementation of \Cref{fig:log1pmd},
    which leverages additional domain knowledge.
}
  but cannot find an accurate implementation of
  either of the other two mathematical expressions in \Cref{fig:somerc},
  illustrating the gap that expert-implemented primitives can fill.

If automated tools are insufficient,
  a numerical methods expert might be the solution.
Typically, a numerical methods expert would
  extract the problematic computation---%
  any of those three expressions above---%
  into a library function and then write
  a custom, accurate implementation of that function,
  guided by their numerical methods expertise.
For example, the expert may choose
  to extract the $\log((1 + x) / (1 - x))$ function
  into a custom helper called, perhaps, \texttt{log1pmd}
  (for ``logarithm, 1 plus minus divide''),
  and then implement that function using, for example,
  a polynomial approximation,
  as in \Cref{fig:log1pmd}.
Or, the expert may be aware of
  an existing implementation of the same function;
  the influential \texttt{fdlibm} math library's implementation of $\log$
  using an implementation of \texttt{log1pmd} as a subroutine,
  and an expert might extract and reuse that subroutine.
The expert's help might be useful even when
  automated tools provide a solution.
For example, the expert implementation in \Cref{fig:log1pmd}
  leverages additional domain information
  about the possible range of $x$
  (it is equal to $e \sin \phi$, where $e$ is
   the eccentricity of the Earth, so $|x| < 0.08182$)
  and is thus substantially faster than Herbie's implementation.

\begin{figure}
\begin{lstlisting}[language=C]
tmerc.cpp:130: xy.x = Q->ml0 * log((1. + b) / (1. - b));
omerc.cpp:62:  v = 0.5 * Q->ArB * log((1. - U) / (1. + U));
isea.cpp:1343: double log1pe_1me = log((1 + P->e) / (1 - P->e));
labrd.cpp:26: V2 = .5 * P->e * Q->A * log((1. + t) / (1. - t));
labrd.cpp:76: V2 = .5 * P->e * Q->A * log((1. + tpe) / (1. - tpe));
labrd.cpp:126: Q->C = .5 * P->e * Q->A * log((1. + t) / (1. - t)) +
somerc.cpp:27: Q->hlf_e * log((1. + sp) / (1. - sp))) +
somerc.cpp:54: Q->hlf_e * log((1. + esp) / (1. - esp))) *
somerc.cpp:89: Q->hlf_e * log((1. + sp) / (1. - sp)));
\end{lstlisting}
\caption{
  Other lines of code in PROJ
    that compute the \texttt{log1pmd} expression.
  Note that some of these uses have the $1 + x$ term
    in the numerator of the division, but others have it
    in the denominator.
  This is common in numerical code
    and means that finding common functions requires
    reasoning about algebraic equivalence and composition.
}
\label{fig:other-uses}
\end{figure}

The problem, of course, is that even a project as influential as PROJ
  cannot expect unlimited attention from numerical methods experts.
They must thus prioritize.
For example, while the Swiss Oblique Mercator
  is surely important for Switzerland,
  it would be a pity to use valuable numerical methods expertise
  on what is essentially a single-country library routine.
A \emph{common, reusable} routine
  that handles a computation common to many projections
  would be more important.
Luckily, \texttt{log1pmd} \emph{is} a common and reusable routine:
  the expression it computes appears across many projections.
But finding these common routines in a large existing codebase,
  to maximally leverage available numerical expertise,
  is not a task any existing tool can assist with.

\name is the first such tool.
\name analyzes a large collection of numerical basic blocks
  and finds common, reusable subroutines
  that a numerical expert can implement
  to improve application accuracy and performance.
These subroutines are short mathematical expressions
  that cannot be improved by current numerical superoptimizers
  yet which appear in many basic blocks
  and significantly influence those blocks'
  accuracy and performance.
As the \texttt{log1pmd} example illustrates,
  recovering reusable primitives from a workload
  poses three challenges:
  recognizing semantically equivalent variants
  across different contexts,
  estimating which candidate abstractions
  would most improve the workload,
  and generalizing beyond the specific expressions
  found in source code.
\name addresses each
  by repurposing existing superoptimizer machinery.

First, \name can reason about semantic equivalence,
  including under algebraic identities.
For example, \Cref{fig:other-uses} shows
  a number of other uses of \texttt{log1pmd} in PROJ.
Many of the matches in \Cref{fig:other-uses} use
  different variable names, or even different expressions, for $x$.
Also, some use $\log((1 + x) / (1 - x))$,
  which is what \texttt{log1pmd} implements,
  while others use $\log((1 - x) / (1 + x))$,
  which would have to call $\mathtt{log1pmd}(-x)$.
\name is nonetheless able to see
  that \texttt{log1pmd} is common to all of them
  because it leverages
  Herbie's e-graph-based equivalence reasoning,
  which can efficiently determine which candidate library functions
  are equivalent or, more weakly,
  can be used to implement one another.

Second, \name can reason about
  the impact of each candidate library function
  on the accuracy and performance of
  the expression as a whole.
For example, \Cref{fig:somerc} also includes
  the expression $e \cdot \sin(\phi)$.
However, this expression is not particularly inaccurate,
  and moreover its performance is less important
  because it is not used along the maximum delay path.
\name thus does not suggest it for implementation.
More generally, \name leverages
  Herbie's accuracy and performance cost model,
  which can estimate counterfactual impact
  (how much a workload would improve
  if a given primitive were available),
  to determine which candidates are worth expert attention.

Finally, \name is able to generalize and abstract
  over the patterns actually found in source code.
For example, in \Cref{fig:somerc},
  inlining the \texttt{sp} variable (as a compiler would do)
  reveals that the actual function being computed
  is not $\log((1 + x) / (1 - x))$
  but the more specific $\log((1 + e\cdot\sin\phi) / (1 - e \cdot \sin \phi))$.
However, this more specific function is \emph{not} common
  to all of the other uses in \Cref{fig:other-uses}.
To generalize and abstract in this way,
  \name leverages Herbie's search procedure,
  which explores algebraic rewrites,
  approximations, and symmetry breaking,
  to find expressions \emph{related} to
  the ones literally in the source code
  that are attractive library functions.

Notably, none of these capabilities are specific
  to \name or to numerical computing.
Equivalence reasoning, impact estimation, and generalization
  are core components of any superoptimizer;
  \name simply repurposes them for library learning
  rather than optimization.
Any domain with a superoptimizer
  (whether for numerical kernels, binary code, SQL queries,
  or hardware circuits)
  could, in principle, apply the same approach
  to discover which new primitives
  would most benefit its workloads.
\Cref{sec:generate,sec:filter} describe
  how \name realizes this connection in detail.

\section{Generating Candidate Primitives}
\label{sec:generate}

This section describes how \name generates
  candidate primitives: subexpressions that,
  if given expert implementations,
  could serve as new primitives
  for the target application's kernels.
The input is the application's kernels,
  represented as basic blocks in FPCore;
  the output is a large, deduplicated set of candidates
  to be ranked by the selection phase (\Cref{sec:filter}).
The core insight is that a superoptimizer's search procedure
  already explores the relevant space:
  to optimize a kernel, it enumerates many equivalent programs,
  and subexpressions of those programs,
  after canonicalization and generalization,
  are natural candidates for new primitives.

Superoptimizers must search through
  a vast collection of programs---ideally, all---%
  that are equivalent to a given input program.
For example, the Herbie numerical compiler / superoptimizer
  uses e-graphs to efficiently search
  through algebraically equivalent terms,
  and also has subcomponents for function approximation
  and symmetry breaking.
It uses this search procedure to select programs
  equivalent to its input program
  but either more accurate or faster to evaluate
  (or some mix of both along a Pareto frontier).

Some of these equivalent programs
  will include novel subsequences of instructions
  with clear semantics that could be extracted
  into a candidate primitive.
Because \name reuses Herbie's search procedure,
  which has been refined over a decade
  of real-world deployment,%
\footnote{
  Herbie has users at national labs (Sandia, NASA)
    and in industry (NVIDIA, Intel),
    and has been the subject of numerous publications
    optimizing its search
    procedures~\cite{herbie,pherbie,chassis,ruler,enumo,egg,egg-proofs,egglog-in-practice,movability,explanifloat}.
}
  it inherits a high-quality, high-throughput
  candidate generation procedure
  without needing to build one from scratch.

More specifically, \name's candidate generation pass
  runs Herbie on all of the input kernels.
For each input kernel,
  Herbie searches over many---%
  typically thousands or tens of thousands---%
  equivalent programs,
  pruning away the vast majority,
  and outputs the best ones it found.
Crucially, \name ignores Herbie's final output,
  which cannot yet benefit from the primitives
  that \name is trying to discover,
  and instead collects all intermediate expressions
  explored \emph{before} pruning.
This inverts the usual role of the superoptimizer:
  rather than using the search to find the best program,
  \name mines the search itself
  for the subexpressions that recur across many kernels.
These intermediates, not the final optimized program,
  are what matter for library learning,
  because they represent the space of expressions
  whose counterfactual utility \name will later estimate.

The programs found during the superoptimizer search
  are equivalent to one of the original kernels.
Since \name is searching for candidate primitives
  that could be \emph{used in} one of those kernels,
  it then considers every subexpression
  of one of the searched programs
  and extracts arbitrary cuts.
For example, if the search step considered the program
  $(\log(1 + x\cdot y))^2$
  then \name would consider subexpressions like
  $\log(1 + x \cdot y)$
  and cuts of such subexpressions like
  $\log(1 + z)$.
To produce cuts, \name does not use a fixed depth.
Instead, it considers cuts
  at every binary or wider operator.
Thus, in $\log(1 + x\cdot y)$,
  the full set of cuts would be $\log(z)$,
  $\log(1 + z)$, and $\log(1 + x\cdot y)$.
However, $\log(z + x \cdot y)$ would not be considered,
  because the constant $1$ is not
  a binary operator.
The intuition is that binary operators increase fan-out
  and thus the number of arguments of candidate primitives
  (primitives with many arguments are harder to implement),
  but unary operators do not increase fan-out
  and there is thus no reason to introduce cuts at them.
In general, for balanced binary trees,
  the number of such cuts grows exponentially,
  but in our experiments this behavior was rare
  and the number of cuts was generally small.
We tested alternative cutting strategies,
  including both simpler fixed-depth strategies
  and also more complex strategies that adapted based on
  the identity of cut operators,
  and found that this simple binary heuristic worked best.

After extracting all cuts,
  \name has a large collection of candidate primitives
  to pass to the selection step.
However, this collection contains many duplicates.
These duplicates happen for two reasons.
First, two expressions might be the same
  except for alpha renaming,
  such as $\log(1 + x)$ and $\log(1 + y)$.
Second, two expressions might be equivalent to each other,
  such as $\log(1 + x)$ and $\log(x + 1)$.
The number of duplicates can be extreme,
  and deduplicating
  makes \name's selection stage much faster.
\name thus deduplicates expressions
  by, first, normalizing all variable names;
  then, considering all permutations of variables;
  and finally using an e-graph,
  again repurposing the superoptimizer's equivalence machinery,
  to prove equivalence between
  as many expressions as possible.
For example,
  given the expression $\frac{x}{1 + y}$, 
  \name first normalizes variable names, 
  producing $\frac{t_1}{1 + t_2}$. 
It then considers all permutations
  of the normalized variables, 
  yielding both $\frac{t_1}{1 + t_2}$ and $\frac{t_2}{1 + t_1}$. 
This allows \name to deduplicate expressions such as $
  \frac{x}{1 + y}$ and $\frac{y}{1 + x}$, 
  which differ only by a renaming of variables.
Any two expressions any of whose permutations were proven equal
  are deduplicated (and the total duplicate count saved for use in selection).
This approach can deduplicate
  expressions like $(1 + x) \cdot y$ and $x \cdot (1 + y)$,
  which can be proven equivalent only by considering
  both alpha renaming (swapping $x$ and $y$)
  and algebraic rearrangement (commutativity of addition).
Nonetheless, this equivalence procedure is expensive
  and requires limiting expressions
  to those with up to three variables.
Since math library primitives with many variables
  are harder to implement,
  this restriction is relatively mild.

\section{Selecting Library Primitives}
\label{sec:filter}

The generation phase produces
  hundreds of thousands of candidate primitives.
The task of the selection phase
  is to choose the small number of these
  that would most improve the target application.
The core insight is that the superoptimizer can directly
  answer the library designer's question:
  if a given primitive existed,
  how much would the workload improve?
This is exactly the counterfactual utility
  that the superoptimizer's cost model
  is designed to estimate.

\subsection{Evaluating Candidates via Platform Extension}

Superoptimizers are generally driven by
  a mechanized semantics for the available instructions
  in the target programming language or platform,
  which are used for proving equivalence to the original program,
  as well as a cost model of some kind
  for evaluating which instruction sequences are best.
Herbie, for example,
  is parameterized by a \emph{platform description}~\cite{chassis}
  which defines all available floating-point operations,
  their input and output types,
  their real-number semantics,
  and functions estimating their cost and accuracy.
Herbie then searches for possible instruction sequences
  using real-number reasoning over instruction semantics
  and combinatorial optimization over their cost functions.
Candidate primitives can then represent
  \emph{extensions} to that platform definition,
  drawing its types and semantics from the primitive's definition
  and adding hypothesized cost and accuracy functions.
Each candidate primitive can be evaluated by
  1) extending the superoptimizer platform with that candidate primitive;
  2) re-optimizing the workload's kernels with this extended platform;
  and 3) measuring the increase in accuracy and performance.
This is the \emph{counterfactual utility} computation:
  the superoptimizer directly answers the question
  ``would this workload be better off
  if this primitive were available?''
We write $\acc{A}{P}$ for the normalized accuracy
  achieved by superoptimizing the kernels in $A$
  when the platform is extended with a set of primitives $P$,
  averaged over all kernels and bounded in $[0, 1]$.

Suppose the target application has a set $K$ of kernels.
The candidate generation step superoptimizes all of $K$,
  yielding a set $C$ of candidate primitives.
\name's ultimate goal is to select a subset $N \subseteq C$
  to maximize $\acc{K}{N}$.
A maximally naive approach,
  enumerating all subsets of $C$ and re-optimizing $K$ with each,
  would require $|K| \cdot |C|^{|N|}$ superoptimizations.
However, this is wildly impractical:
  in our evaluation we have roughly
  $|K| \approx 10^2$, $|C| \approx 10^6$, and $|N| \approx 10^1$,
  meaning this would take approximately $10^{62}$ superoptimizations,
  which each take a few seconds.
\name thus applies a series of heuristics and optimizations
  to reduce the number of superoptimizations required
  to some reasonable number.

We discuss the ranking heuristics below
  but first focus on another heuristic:
  the cost and accuracy functions used
  when extending the superoptimizer platform with a candidate primitive.
It is critical that the cost and accuracy definitions be \emph{optimistic}:
  the intention is for a human expert
  to help implement the selected candidate primitives,
  so their cost and accuracy will be \emph{better}
  than the implementations found by the superoptimizer.
To this end, we tell the superoptimizer to assume that candidates
  can be implemented with no error and a cost that is
  $1/5$ of the cost of their naive implementation.
These optimistic cost and accuracy definitions
  will incentivize the superoptimizer
  to find uses for the candidate primitives,
  which will help determine which candidate primitives are most useful.
There is, of course, a risk of being \emph{over}-optimistic,
  in which case the selected candidates may simply be
  impossible to implement efficiently,
  a problem we discuss further in \Cref{sec:implementation}.
These optimistic assumptions are used only for ranking;
  \Cref{sec:eval} validates the final selections
  using real expert implementations.
To reduce this risk, however,
  the ranking heuristics favor terms that are
  more likely to be efficiently implementable.

\subsection{Ranking Candidates by Size, Frequency, and Urgency}

The first simplification \name makes is to
  use a greedy strategy to grow the set of selected primitives
  one primitive at a time.
That is, \name maintains a set
  of already-selected candidate primitives $S$.
It then considers each non-selected candidate primitive $f$
  and searches for one that maximizes $\acc{K}{S \cup \{f\}}$,
  adding it to $S$ and repeating the process
  until $S$ is the necessary size.
A naive implementation of such a greedy algorithm,
  however, would still require $|K| \cdot |C| \cdot |N|$ superoptimizations,
  still too large at around $10^9$.

To further reduce runtime, \name observes that
  a candidate primitive must have
  three properties to be worth selecting,
  each of which can be estimated
  from the superoptimizer's existing machinery.
First, a size heuristic is based on the candidate's cost in 
  the superoptimizer's cost model.
This heuristic favors smaller candidate primitives,
  meaning a smaller AST describing their real-number behavior.
Smaller candidate primitives are typically easier
  for a person to understand and implement,
  and they're less likely to be overly specialized.
Second, a frequency heuristic favors candidate primitives
  that can be used more frequently in the target application.
Frequency counts how often the superoptimizer
  considered an expression containing the candidate primitive.
This exploits the superoptimizer's search procedure
  as a proxy for how likely the superoptimizer was
  to use the candidate primitive, if it were available.
Third, an urgency heuristic favors candidate primitives
  that the superoptimizer is unable to improve on its own.
If the superoptimizer can find a high-quality implementation
  of the candidate primitive on its own,
  then there is no need for an expert to write their own.
Urgency is measured by $1 - \acc{f}{S}$,
  meaning by superoptimizing each candidate $f$
  given the already-selected candidates $S$;
  accuracy is then given by the gap between
  actually-achieved accuracy and maximum accuracy.
The three heuristics are combined in a straightforward way,
  taking the product of frequency and urgency
  and dividing by size.
We've observed that candidate primitives
  that score highly on the three heuristics
  are also those most likely to increase \acc{K}{S}.
Moreover, the heuristics are cheaper to measure:
  size and frequency can be measured during
  the original generation stage,
  while the urgency heuristic just requires $|N| \cdot |C|$ superoptimizations.
Adding $|K|$ superoptimizations for the generation stage,
  this brings the runtime down to $|K| + |N| \cdot |C| \approx 10^7$ superoptimizations.

In practice,
  all three heuristics help separate
  promising candidates from unpromising ones.
That said, low-ranked candidates by one measure
  are also often bad by some other measure.
For example, a long, overly-complex candidate primitive
  is likely to be not only large (failing the size heuristic)
  but also overly specialized (failing the frequency heuristic),
  while a candidate primitive
  with many unnecessary or duplicative operations
  is likely to be both overly large (failing the size heuristic)
  and optimizable through simple means (failing the urgency heuristic).
This suggests filtering first on the size and frequency heuristics,
  which do not require superoptimizations to evaluate,
  and only using urgency as a second pass.
We therefore split the heuristic ranking into two stages:
  we first select the top $T_1$ candidate primitives
  by the size and frequency heuristics,
  and then rank just those on urgency.
This reduces runtime from $|K| + |N| \cdot |C|$ superoptimizations
  to $|K| + |N| \cdot T_1$; \name sets $T_1 = 625$,
  so this is approximately $6 \times 10^3$ superoptimizations.
Setting $T_1$ high ensures that the overall ranking
  isn't changed much:
  the top-ranked candidates by size, frequency, and urgency
  must be among the top $T_1$ by size and frequency alone
  because relatively many candidates (roughly $4\%$ in our evaluation)
  have high urgency.
Candidates with low size and frequency could not
  have high enough urgency (since urgency is bounded in $[0, 1]$)
  to end up higher ranked than them.

\begin{figure}
\includegraphics[width=0.33\columnwidth]{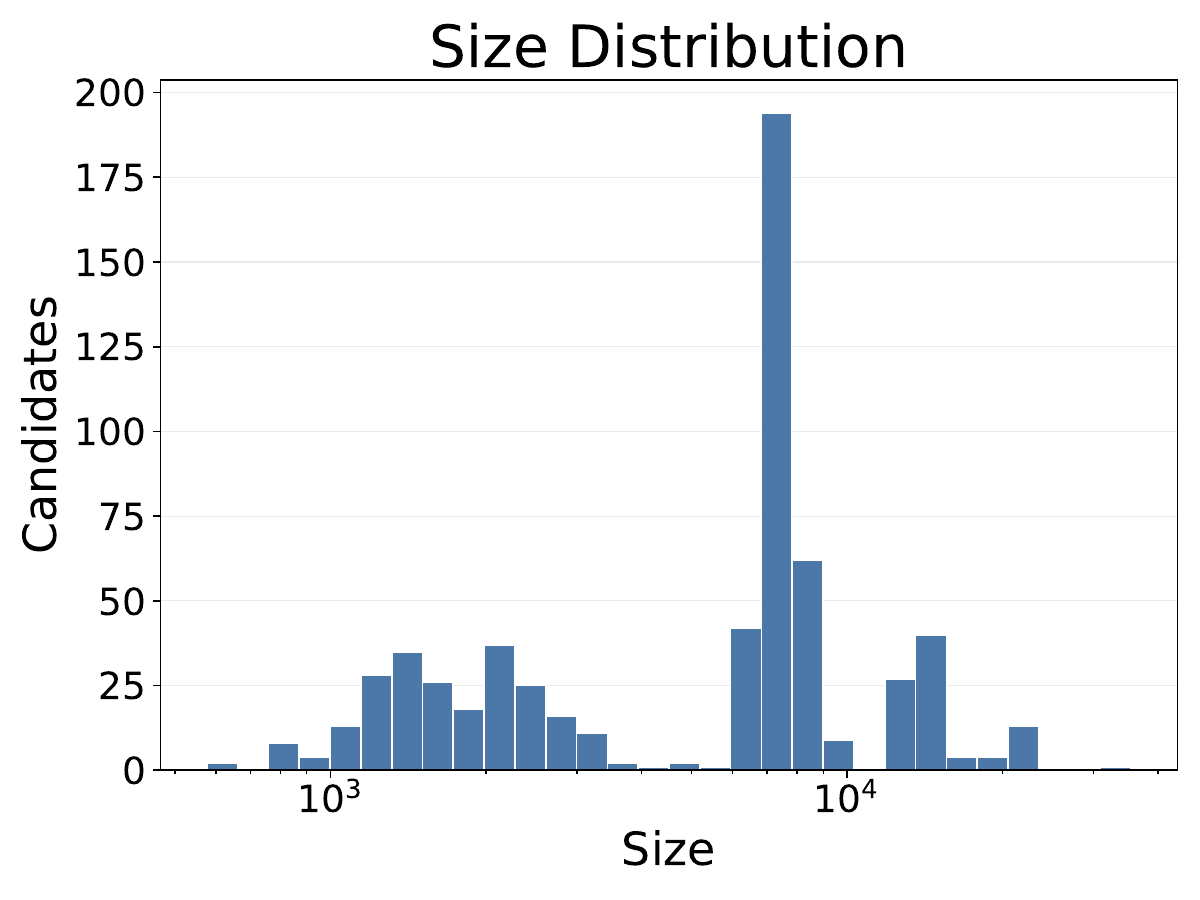}%
\includegraphics[width=0.33\columnwidth]{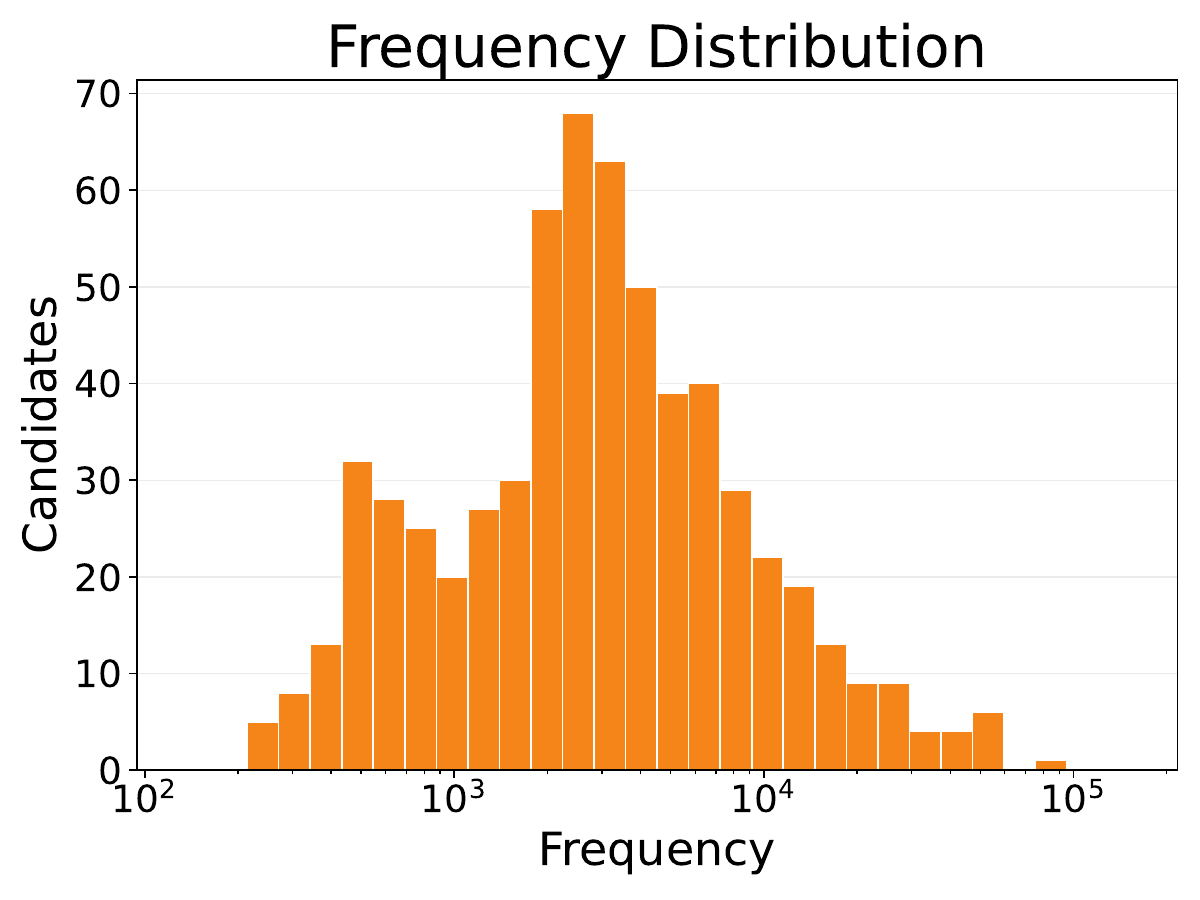}%
\includegraphics[width=0.33\columnwidth]{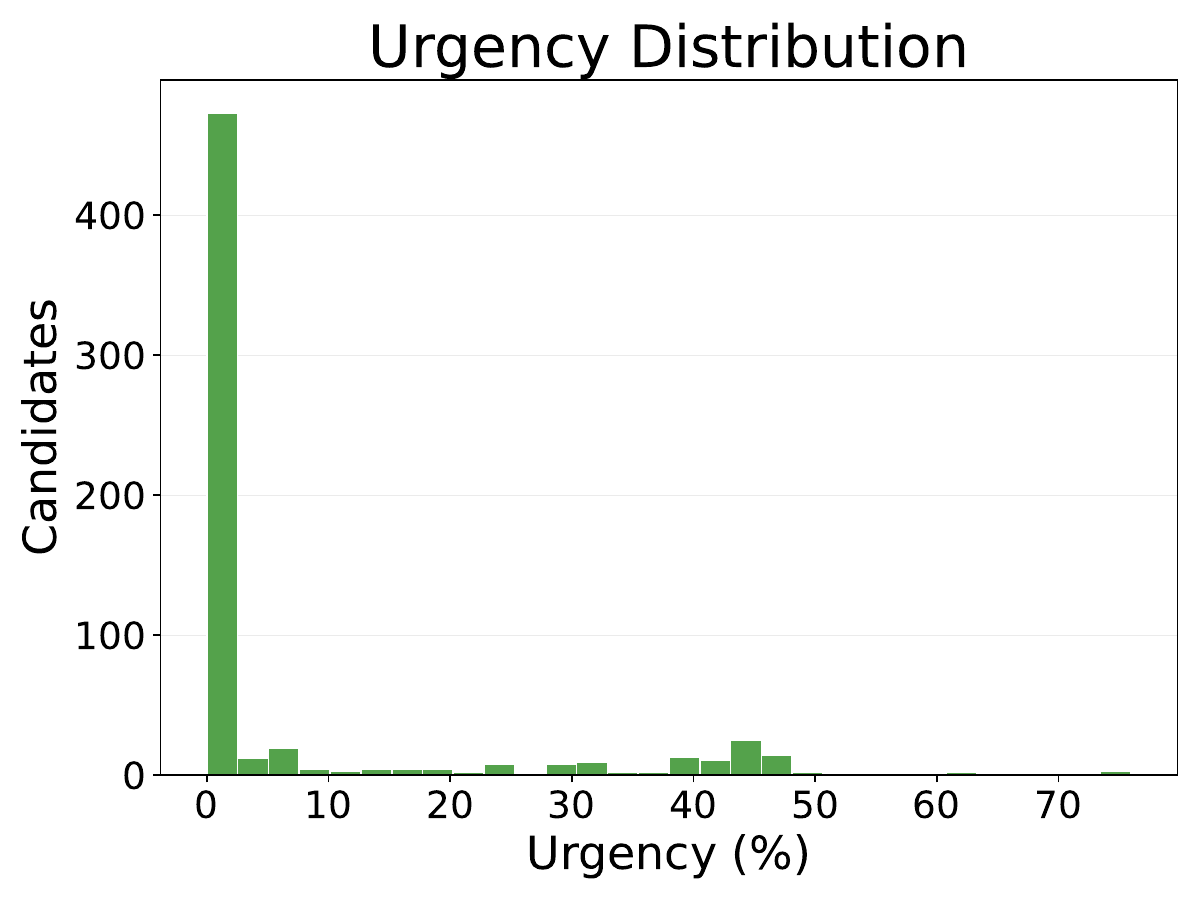}%
\caption{
  Distributions of the size, frequency, and urgency
    heuristics for a PROJ run.
  Size and frequency are well-distributed,
    providing good discriminative signal.
  Urgency is concentrated near zero
    with a long tail of high-urgency candidates,
    confirming the bimodal structure
    that justifies two-stage filtering.
}
\label{fig:histograms}
\end{figure}

\Cref{fig:histograms} confirms this structure empirically:
  size and frequency spread across candidates,
  while urgency is sharply bimodal,
  with roughly $96\%$ of candidates near zero
  and a small fraction with high urgency.
This is precisely the distribution
  that makes filtering by size and frequency first,
  then ranking on urgency, effective.

\subsection{Pruning Redundant Candidates via Implication}
To reduce runtime further,
  \name selects candidates in batches
  rather than one at a time.
However, simply selecting the top few candidates
  by size, frequency, and urgency
  often selected variants of the \emph{same} candidate.
For example, both $\log(1 + x)$ and $\log(1 + x \cdot y)$
  might rank highly on size, frequency, and urgency.
These candidate primitives are not equivalent,
  so they will not be deduplicated in the generation phase.
Yet any use of $\log(1 + x)$ can be replaced
  by calling $\log(1 + x \cdot y)$ with $y = 1$;
  conversely, any use of $\log(1 + x \cdot y)$ can be replaced
  by calling $\log(1 + x)$ with its argument set to $x \cdot y$.
Asking a human expert to implement both would be a waste of their time.
In fact, seeing variants like this is expected:
  these candidates often had similar frequency
  because they were both cuts of the same initial expression
  and similar urgency due to similar numerical problems
  (here, the zero that $\log$ has at $1$).
For another example, consider the candidate primitive
  $\operatorname{pow}(\cos(x), 6)$;
  any use of this candidate primitive can be replaced
  by a use of the more general $\operatorname{pow}(x, 6)$ function,
  but the reverse is not true, since $\cos(x)$ does not range over
  the full domain of $x$.
In this case,
  it would be better to select the more general candidate primitive
  and defer selecting the more specific one
  until its urgency could be recomputed.

Resolving such implications is critical
  to adding multiple top-ranked candidates at a time.
Consider a ``batch'' of $T_2$ top-ranked candidates
  by size, frequency, and urgency.
\name measures \acc{g}{S \cup \{f\}}
  for each pair of candidates $f$ and $g$ in the batch;
  if the result passes 95\%,
  it records that $f \prec g$,
  meaning that selecting $f$ makes $g$ redundant.
Since the batch already contained candidates
  that rank highly on urgency,
  \acc{g}{S} is generally low;
  if \acc{g}{S \cup \{f\}} is high,
  then selecting $f$ obviates the need to select $g$.
This reuses the counterfactual utility computation:
  the superoptimizer determines whether one candidate
  renders another redundant by checking whether
  the workload improves just as much without it.
The $\prec$ relation forms a directed graph
  of the $T_2$ candidate primitives in the batch;
  \name partitions this graph into a directed acyclic graph
  of strongly-connected components~\cite{scc}
  and selects one candidate primitive
  from each strongly connected component
  with no incoming edges in the acyclic portion of the graph.
This ensures that none of the selected primitives
  obviate each other, avoiding wasted expert work.
Assuming a constant fraction $\alpha$
  of the $T_2$ candidate primitives is selected,
  the number of superoptimizations needed is reduced
  to $|K| + (|N| / \alpha T_2) (T_1 + T_2^2)$.
In \name, $T_2$ is set to $25$
  and $\alpha$ is approximately $\frac14$,
  reducing the number of superoptimizations to
  roughly $10^2 + \frac{10}{0.25 \cdot 25} (625 + 25^2) \approx 2{,}100$.
Since each superoptimization takes a few seconds,
  the final pipeline runs in a few hours
  (see \Cref{sec:eval} for more).

\paragraph{Tuning $T_1$ and $T_2$ for tractable selection.}

Both $T_1$ and $T_2$ are parameters
  that trade off between the quality of the selected primitives
  and the time taken by the selection phase.
To select $T_1$, we observed that our average urgency
  across candidate primitives was roughly $4\%$.
In practice, the urgency is very bimodal---%
  roughly $96\%$ of candidates have barely any urgency
  and the remaining roughly $4\%$ have high urgency.
Thus, the top $.04 T_1$ candidates
  by the size, frequency, and urgency heuristics
  are likely to be among the top $T_1$ candidates
  by just size and frequency.
Since the implication step only uses a batch
  of the top $T_2$ candidates,
  this suggests that $T_2 = .04 T_1$.
At the same time, to minimize the number of superoptimizations,
  $T_1$ and $T_2$ should be chosen to minimize
  $(1 / T_2) (T_1 + T_2^2)$.
This function is minimized at $T_2 = \sqrt{T_1}$.
Combining both formulas yields
  $T_1 = 625$ and $T_2 = 25$.

\section{Implementing Library Primitives}
\label{sec:implementation}

\name suggests candidate primitives,
  but does not implement them automatically;
  that task requires numerical expertise.
A key question, then, is whether \name's suggestions
  are at the right scope for bounded expert effort.
To test this, our evaluation required implementing
  each \name-selected primitive.
This section details those implementations
  and demonstrates that a range of standard techniques
  (existing library functions, error-free transformations,
   range reduction, and polynomial approximation)
  suffice to produce fast, accurate implementations.
For each primitive, we aimed for 1-ULP or few-ULP accuracy
  while keeping cost low enough for Herbie to prefer them.
To aid readers (and ourselves),
  we assign names to each primitive below,
  though \name itself outputs only mathematical specifications.
Unless otherwise noted, all primitives below are from PROJ,
  the largest and most diverse of our three workloads
  (92 kernels across five projections);
  CoolProp and Basilisk each produced
  a smaller, more focused set.

Our implementations were guided by one of the authors,
  an expert on elementary function implementation.
However, to bound the effort available for implementations,
  we asked the expert to only communicate short instructions
  to another author, a student
  who had never previously implemented math functions;
  the expert author performed no implementation on their own.
Besides their own abilities
  and the assistance of GPT~5.3 Codex,
  the second author additionally used components from
  the well-known \texttt{fdlibm} library~\cite{fdlibm}
  as well as the Sollya tool~\cite{sollya}
  for generating polynomial approximations,
  both suggested by our human expert.
Each implementation was done in C,
  compiled by Apple clang version 17.0.0,
  and executed in a timing and testing harness.
The harness compared the computed results to
  a high-precision result produced by the Rival library~\cite{movability}
  to estimate accuracy
  and also timed the executions to estimate cost.
The resulting accuracies and costs were then used
  to create a new Herbie platform with
  all candidate primitive implementations,
  which was used to again superoptimize the target application.
  
\subsection{Existing Library Functions}

Some of \name's suggested primitives
  already have implementations in standard math libraries
  or can be handled by Herbie with higher tuning parameters,
  validating that \name's selections
  align with known numerical practice.

\name selects the standard primitive
  $\mathsf{hypot}(x, y) = \sqrt{x^2 + y^2}$.
This function is already available in standard math libraries,
  so no additional effort was needed to implement it
  besides linking to the correct function name.

\name selected the primitive
  $\mathsf{verdcos}(x) = \cos(2 x) - 1$,
  a variant of what's known as the ``vercosine'' function.
While Herbie could not find a better implementation
  at the standard settings used by \name,
  simply raising tuning parameters like the number of iterations
  allowed Herbie to propose the implementation
  $\mathsf{verdcos}(x) = -2 \sin(x)^2$,
  which is fast and accurate.
Presumably,
  if \name had used these higher parameters
  in its generation and selection stages,
  \textsf{verdcos} would not have been selected,
  as it would have had low urgency.

\name selected the primitive
  $\mathsf{log1pmd}(x) = \log((1 + x) / (1 - x))$.
Running Herbie with higher tuning parameters
  allowed Herbie to propose the implementation
  $\mathsf{log1pmd}(x) = 2 \mathsf{atanh}(x)$,
  but we also noticed that \texttt{fdlibm}
  includes an optimized implementation
  using a degree-7 polynomial,
  which was faster and equally accurate.

\subsection{Double-double tricks}

Other primitives require precision-extending techniques
  to handle rounding error in their inputs.

\name selected the primitive
  $\mathsf{sinprod}(x, y) = \sin(x \cdot y)$,
  and an analogous cosine product primitive.
When $x$ and $y$ are large,
  relative to the period $2 \pi$ of $\sin$,
  the direct implementation is imprecise
  because the rounding error can be larger
  than the period.
Our human expert proposed computing
  $x \cdot y$ as an ``error-free transformation''~\cite{eft},
  which returns both
  the typical floating-point product $z = x \hat{\times} y$
  and its error $e = \mathsf{fma}(x, y, -z)$;
  then $\sin(z + e)$ is evaluated using
  the sine sum-of-angles identity.
For maximum performance,
  the standard library \textsf{sincos} function was used,
  which computes both $\sin(z)$ and $\cos(z)$
  in a single, faster function call,
  and likewise for $\sin(e)$ and $\cos(e)$.

\name also selected the primitive
  $\mathsf{sinquot}(x, y) = \sin(x / y)$,
  and the analogous cosine quotient primitive.
An initial implementation
  based on error-free transformations, analogous to \textsf{sinprod},
  wasn't particularly accurate,
  which our human expert suggested was due to the fact that
  the error-free transformation for $x \cdot y$ is exact,
  while the one for $x / y$ is not.
Our human expert then suggested an implementation where,
  for large values of $x$,
  the $y$ parameter was split into two halves $y_h$ and $y_l$,
  and then Payne-Hanek reduction was used
  to compute $x \bmod y\pi/2$
  using a high-precision multiplication of $\pi/2$ with $y_h$ and $y_l$.
This did produce a highly-accurate implementation
  (via the efforts of GPT~5.3 Codex,
   with minimal additional prompting),
  but the resulting implementation was exceptionally slow,
  and Herbie did not choose to use it in
  any target application basic blocks.
Our expert suggested that
  further performance improvements would be difficult.

\subsection{Range Reduction and Polynomial Approximation}

The most common implementation pattern combines
  range reduction with polynomial approximation,
  a standard technique in math library development.

\name selected the primitive
  $\mathsf{invgud}(x) = \log(\tan(\pi/4 + x/2))$.
This function is apparently called the ``inverse Gudermannian''
  and is related to the Mercator map projection.
Our human expert suggested a range reduction
  using the \texttt{rem\_pio2} function from \texttt{fdlibm}
  followed by a Sollya-generated polynomial approximation.
This produced a highly-accurate implementation
  (1 ULP of error on all tested points)
  that was also highly efficient.

For the Basilisk workload,
  \name selected the primitives
  $\mathsf{powcos2}(x) = \mathsf{pow}(\cos(x), 2)$,
  $\mathsf{powcos4}(x)$, and $\mathsf{powcos6}(x)$.
For these functions,
  our initial attempt used the \texttt{rem\_pio2} range reduction
  and a Sollya-generated polynomial approximation,
  but the results were too slow
  and Herbie did not choose to use them.
Our expert suggested a faster range reduction
  from the CORE-Math library~\cite{core-math},
  which was substantially faster though less accurate on a few points.
This faster implementation was in fact used by Herbie.

For CoolProp, \name selected the primitive
  $\mathsf{pow1ms}(x, y) = \mathsf{pow}((x - 1)^2, y)$.
Our expert suggested rewriting this as
\[
  \mathsf{pow}((x - 1)^2, y) 
  = \exp(\log((x - 1)^2) \cdot y)
  = \exp(\log(1 + x\cdot(x - 2)) \cdot y)
\]
  and using the standard \textsf{log1p} function
  to evaluate $\log(1 + z)$, with $z = x\cdot(x - 2)$.
This implementation was fast enough to be used by Herbie.

In short, standard techniques sufficed
  for all but two of \name's suggested primitives.
The exceptions, \textsf{sinquot} and \textsf{cosquot},
  demanded multi-word arithmetic
  that our expert judged too slow to be practical.
The remaining primitives were all implemented
  to near-1-ULP accuracy
  without requiring novel numerical methods,
  suggesting that \name tends to select primitives
  at a useful level of difficulty.

\section{Evaluation} \label{sec:eval}

We evaluate \name's suggested primitives
  across six research questions:

\begin{enumerate}
\item[RQ1] What new primitives does \name propose?

\item[RQ2] Can an expert implement these primitives well?
  
\item[RQ3] Can Herbie use the new primitives
  to optimize the targeted kernels?

\item[RQ4] Do the new primitives
  improve performance and error
  of the targeted kernels?

\item[RQ5] Can learned primitives be deployed
  in compiled applications?

\item[RQ6] How long does \name take to run?
\end{enumerate}

\noindent
We evaluate \name on 3 open-source target applications
  from different scientific domains:

\begin{enumerate}
\item PROJ~\cite{proj} is a library
  of geodetic transformations 
  and cartographic projections,
  for example for transforming geospatial datasets
  from one coordinate reference system to another.

\item CoolProp~\cite{coolprop} is a library
  for calculating the thermophysical properties of fluids,
  for example for modeling steam turbine power systems.

\item Basilisk, an astrodynamics simulation framework 
  used for modeling spacecraft dynamics,
  for example for orbit planning or guidance system development.
\end{enumerate}

Each project is large enough
  that superoptimizing the full library with Herbie is infeasible.
Instead, we selected a subset of each project:
  we extracted 92 kernels from PROJ
  (from five files implementing the
   Krovak, Oblique Mercator, Transverse Mercator,
   Space Oblique Mercator, and Swiss Oblique Mercator projections),
  24 kernels from CoolProp
  (from one file implementing Helmholtz energy formulations)
  and 28 kernels from Basilisk
  (from one file implementing orbital mechanics calculations).
We chose related kernels within each project
  because library learning targets common, reusable primitives,
  which are more likely to emerge from kernels
  in the same application domain.
We also restricted ourselves to kernels
  that call at least one elementary function,
  since those are the kernels that Herbie targets.
We converted each kernel into \name's input format (FPCore~\cite{fpcore,fpbench})
  and then used each project as a separate target application for \name.
We perform all experiments except RQ5 (see \Cref{subsec:RQ5})
  on a machine with an Intel i7-8700K CPU and 32\,GiB of RAM
  running Ubuntu~24.04.4, Rival~2.3, Egg~0.11.0, and Racket~9.1.

The following subsections trace \name's suggestions
  from proposal through deployment.
RQ1--RQ3 use \Cref{fig:eval:function-table}
  to show that \name proposes compact, domain-appropriate primitives,
  that these can be implemented accurately
  (we report Herbie's average bits of error; lower is better)
  and efficiently
  (we report inverse throughput relative to a baseline; lower is better),
  and that Herbie reuses them extensively.
RQ4 then asks whether these implementations
  actually improve kernel-level performance,
  using speed/accuracy frontier curves (\Cref{fig:speedup})
  and per-kernel error comparisons (\Cref{fig:accuracies}).
RQ5 tests deployment in compiled PROJ code via a prototype LLVM matcher,
  and RQ6 reports \name's own runtime.

\subsection{RQ1: What Primitives Does \name Propose?}

\name mines primitives from the superoptimizer's search space.
Are the results compact, recognizable, and domain-appropriate?
The first two columns of \Cref{fig:eval:function-table}
  list all primitives selected by \name
  for PROJ, CoolProp, and Basilisk (in order).

\begin{figure}
\begin{tabular}{ll | rr | rr | r}
& & \multicolumn{2}{|c|}{Naive} & \multicolumn{2}{|c|}{Expert} & \\
Name & Formula & Cost & Error & Cost & Error & Uses\\\hline

hypot & $\sqrt{x^2 + y^2}$ & 1.201 & 29.571 & \bf 0.864 & \bf 0.000 & 39\\
invgud & $\log\left(\tan\left((2x + \pi)/4\right)\right)$ & 4.186 & 59.875 & 9.955 & \bf 0.082 & 22\\
log1pmd & $\log\left((1 + x)/(1 - x)\right)$  & 2.437 & 58.944 & \bf1.093 & \bf0.000 & 30 \\
cosprod & $\cos(xy)$ & 2.106 & 27.309 & 4.086 & \bf 7.684 & 4\\
sinprod & $\sin(xy)$ & 2.145 & 27.940 & 4.086 & \bf 7.728 & 9\\
cosquot & $\cos(x/y)$ & 2.192 & 26.896 & $1.9\cdot10^6$ & \bf 0.100 & 0\\
sinquot & $\sin(x/y)$ & 2.231 & 28.250 & $1.9\cdot10^6$ & \bf 0.239 & 0\\
verdcos & $\cos(2x) - 1$ & 2.441 & 15.306 & 2.562 & \bf 0.272 & 10\\\hline

pow1ms & $\pow((x - 1)^2, y)$ & 3.311 & 15.099 & \bf 2.855 & \bf  0.041 & 87\\\hline

powcos2 & $\cos(x)^2$  & 4.104 & 0.285 & \bf 4.056 & 0.355 & 2\\
powcos4 & $\cos(x)^4$  & 4.104 & 0.414 & \bf 3.461 & 0.449 & 41\\
powcos6 & $\cos(x)^6$  & 4.104 & 0.628 & \bf 3.884 & 0.726 & 6\\
\end{tabular}

\caption{
  Library primitives suggested by \name
    on three target applications:
    PROJ, CoolProp, and Basilisk
    (in order, separated by horizontal lines).
  For each library primitive, ``Naive Implementation'' shows
    the cost (inverse throughput, relative to a baseline)
    and error (Herbie's average bits of error metric)
    of implementing the formula directly,
    while ``Expert Implementation'' shows the cost and error
    of the implementation suggested by our expert.
  \name's suggestions are recognizable as
    simple, generic primitives
    whose naive implementations are typically inaccurate,
    while the expert implementations are, by and large,
    extremely accurate with acceptable costs.
}
\label{fig:eval:function-table}
\end{figure}

All have short, generic specifications.
Some, like \texttt{hypot}, are available in standard math libraries.
Others, like \texttt{invgud} and \texttt{verdcos},
  are (or are variations of) well-known in their domain.
For example, \texttt{invgud} is the inverse Gudermannian function ,
  which is known to be related to Mercator projections~\cite{wiki-invgud}.
Some functions are surprising;
  for example,
  we were surprised when Herbie rated
  the \texttt{cosprod} and \texttt{sinprod} primitives
  as inaccurate when implemented naively.
However, consultation with our expert revealed
  that for very large inputs $x$ and $y$ (such that $x y > 2^{52}$),
  the rounding error of $x y$ can be comparable to or bigger
  than the period of $\sin$ or $\cos$,
  in which case more sophisticated implementation
  of these primitives can improve accuracy.
Similarly, we were initially confused about
  why the \texttt{pow1ms} function squared the $x - 1$ term.
Our expert explained to us that,
  because $(x - 1)^2$ is guaranteed to be positive,
  this form of \texttt{pow1ms} is easier to implement accurately;
  negative first arguments for \pow are apparently difficult to handle.
In short, examination of \name's proposed primitives
  supports the idea that these are well-chosen, useful primitives
  for each target application's respective domain.

Moreover, note that the three target applications
  have wholly-disjoint sets of \name-suggested primitives.
This shows that \name is
  suggesting primitives specific to each target.
This is further supported by, for example,
  the inverse Gudermannian function being known
  to be related to Mercator projections,
  variants of which were the source of
  many of our PROJ kernels.

\subsection{RQ2: Can Experts Implement \name-selected Primitives?}

For numerical library learning to be practical,
  the suggested primitives must be at the right scope
  for bounded expert effort.

The next four columns of \Cref{fig:eval:function-table}
  list the cost and error of both
  the naive and expert implementations
  of each \name-suggested primitive.
Lower cost and error is better;
  the naive implementation is always a direct translation
  of the formula to floating-point operations,
  while the expert implementation is that
  described in \Cref{sec:implementation}.
Error is measured by Herbie
  and cost is measured by a small benchmarking harness;
  both use sampling over thousands of inputs.
The expert-written implementations are generally highly accurate;
  other than \texttt{sinprod} and \texttt{cosprod},
  all have less than 1 bit of average error.%
\footnote{
Our expert implementations of \texttt{sinprod} and \texttt{cosprod}
  are inaccurate on points where $x y$ overflows,
  where both our expert implementation
  and the naive implementation return NaN.
The expert implementations are still
  much more accurate than a naive implementation
  on all other points.
Our expert did suggest a method to implement
  \texttt{sinprod} and \texttt{cosprod} accurately
  even on these most challenging inputs
  but we didn't have time to do so.
}
The naive implementations \emph{are} generally faster,
  but their inaccuracy makes them nearly unusable,
  and the expert-written accurate versions
  are typically only a few times slower
  than the naive versions.
For example, while the expert-written \texttt{invgud}
  is more than four times slower
  than a naive implementation,
  the naive implementation is measured to have
  almost 60 bits of error by Herbie,%
\footnote{
In Herbie's metric, this means the naive implementation
  frequently has the incorrect \emph{exponent}.
}
  while the expert implementation has
  less than 0.1 bits of error.
Three expert implementations---%
  \texttt{hypot}, \texttt{log1pmd}, and \texttt{pow1ms}---%
  are both faster \emph{and} more accurate
  than a naive implementation.
On the other hand, the three \texttt{powcos} implementations
  are already accurate in their naive implementation,
  but their expert implementation is significantly faster.
In short, most \name-suggested primitives
  were implemented accurately and efficiently by an expert.

The only substantial exception to these good results
  are the \texttt{cosquot} and \texttt{sinquot} primitives.
These primitives are challenging to implement accurately
  because, unlike \texttt{cosprod} and \texttt{sinprod},
  the rounding error of a division operation
  cannot be captured exactly as a floating-point number.
The expert-recommended implementation thus requires
  extremely expensive multi-word arithmetic operations and,
  while the result \emph{is} extremely accurate,
  it is far too slow for practical use.
Further consultation with our expert suggested that,
  while the implementation could potentially be sped up somewhat,
  our expert did not see a path toward making these implementations
  accurate enough and fast enough
  for real-world practical use.%
\footnote{Our expert did describe
  the task of implementing these two primitives accurately
  as a ``fun puzzle''.}

\subsection{RQ3: Does Herbie Use the \name-suggested Primitives?}
\label{subsec:RQ3}

This research question directly tests \name's
  counterfactual utility predictions:
  if \name predicted that a primitive would be useful,
  does Herbie actually use it when the primitive is available?

The final column of \Cref{fig:eval:function-table}
  lists how often Herbie used each \name-suggested primitive
  to optimize each target application's kernels.
More specifically,
  we wrote three new platform definition files for Herbie,
  one for each target application,
  listing all \name-suggested primitives \emph{for that application},
  including their cost and accuracy functions.
All \name-suggested primitives are used,
  except for \texttt{cosquot} and \texttt{sinquot}
  (which, as discussed above, are too slow for practical use).
In sum, all target applications use the
  \name-suggested primitives dozens of times,
  suggesting that these expert-written implementations
  are genuinely useful in implementing kernels accurately and efficiently.
Most notable is CoolProp.
Despite the fact that \name only suggests
  a single primitive for CoolProp, \texttt{pow1ms},
  this one primitive is used 87 times across the CoolProp kernels,
  suggesting that it is so useful that many kernels invoke it multiple times.
This is likely because this primitive's expert implementation
  is both faster and more accurate than a naive implementation.
In short, expert implementations of \name-suggested library primitives
  are useful in superoptimizing all of our target applications.

\subsection{RQ4: Do \name-suggested Primitives Improve Kernel Performance?}
\label{subsec:RQ4}

\begin{figure}

\includegraphics[width=0.33\columnwidth]{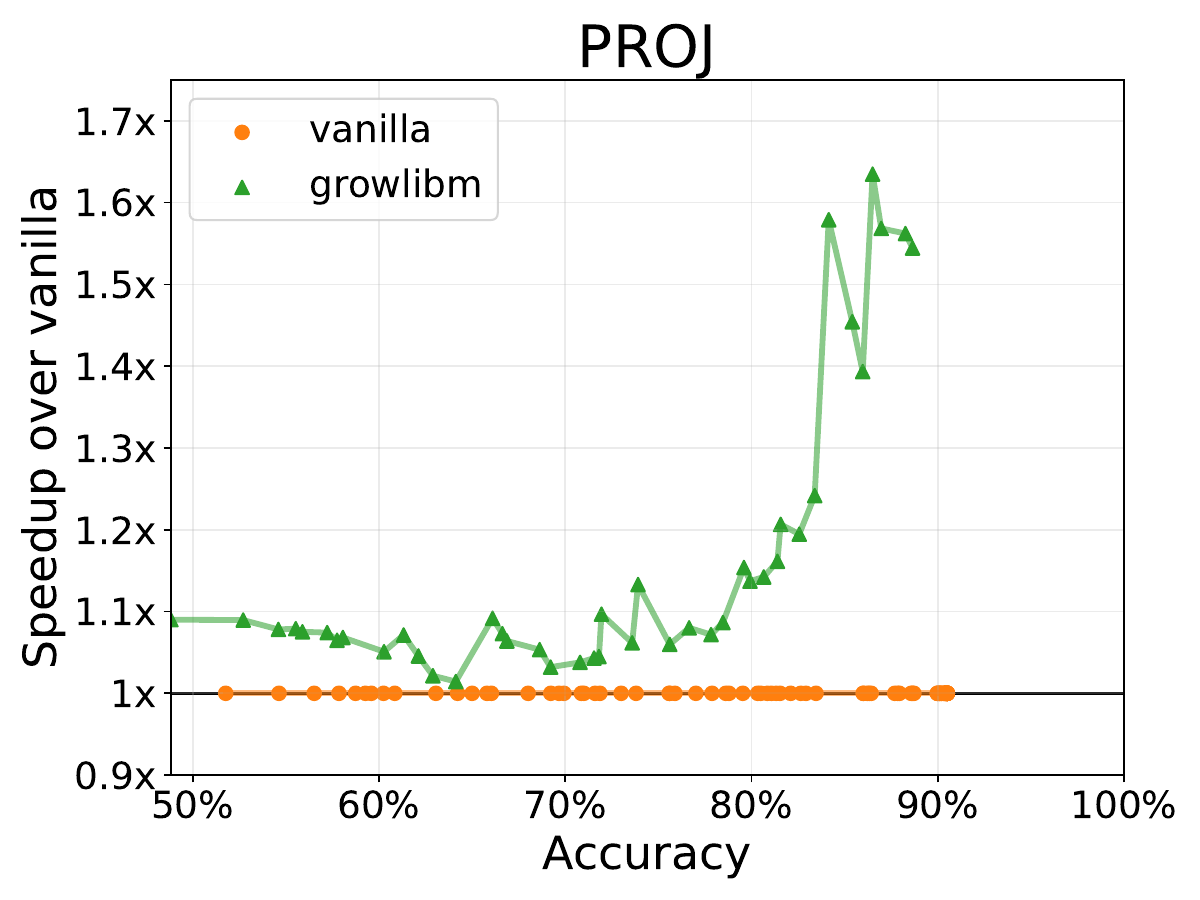}%
\includegraphics[width=0.33\columnwidth]{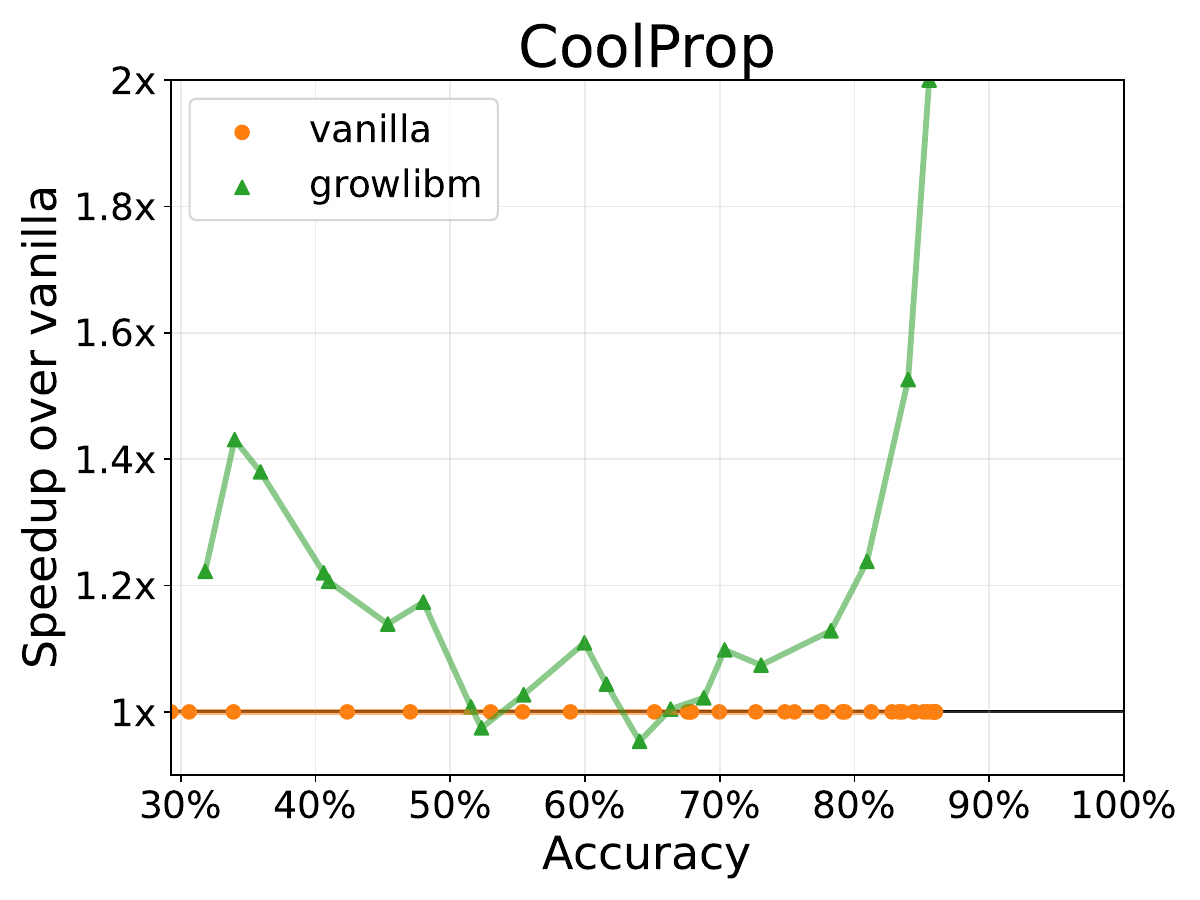}%
\includegraphics[width=0.33\columnwidth]{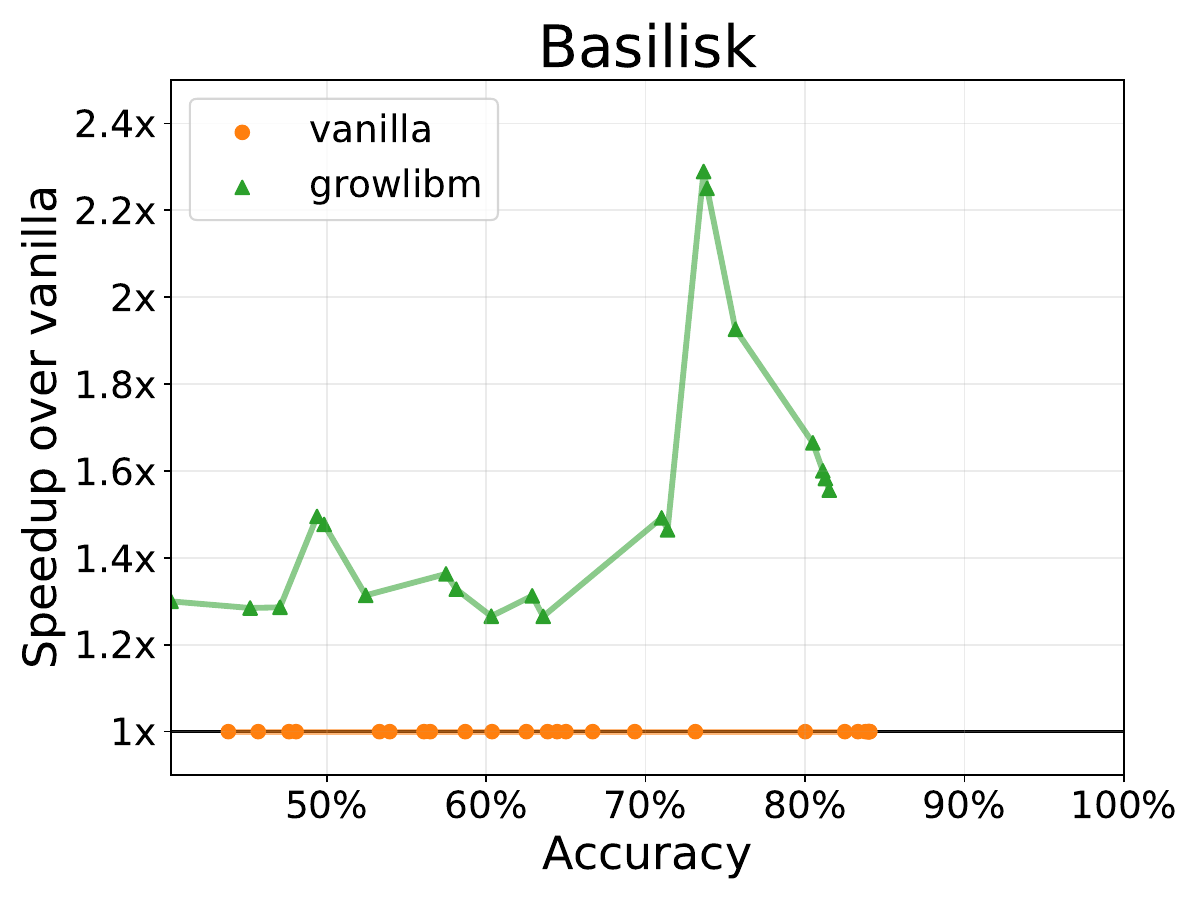}
\caption{
  Speedup curves for PROJ, CoolProp, and Basilisk;
    higher is better.
  Each curve compares Herbie cost estimates
    with and without the \name-suggested library primitives
    for a given level of accuracy.
  Specifically, each curve aggregates
    all of the kernels in the target application
    where a \name-suggested library primitive is used
    and which meet a given average accuracy level.
  \name-suggested library primitives
    significantly speed up implementations
    across the accuracy spectrum,
    with the largest speedups at high accuracies
    where expert implementations of \name-suggested primitives
    are most important.
}
\label{fig:speedup}
\end{figure}

\begin{figure}
\includegraphics[width=\columnwidth]{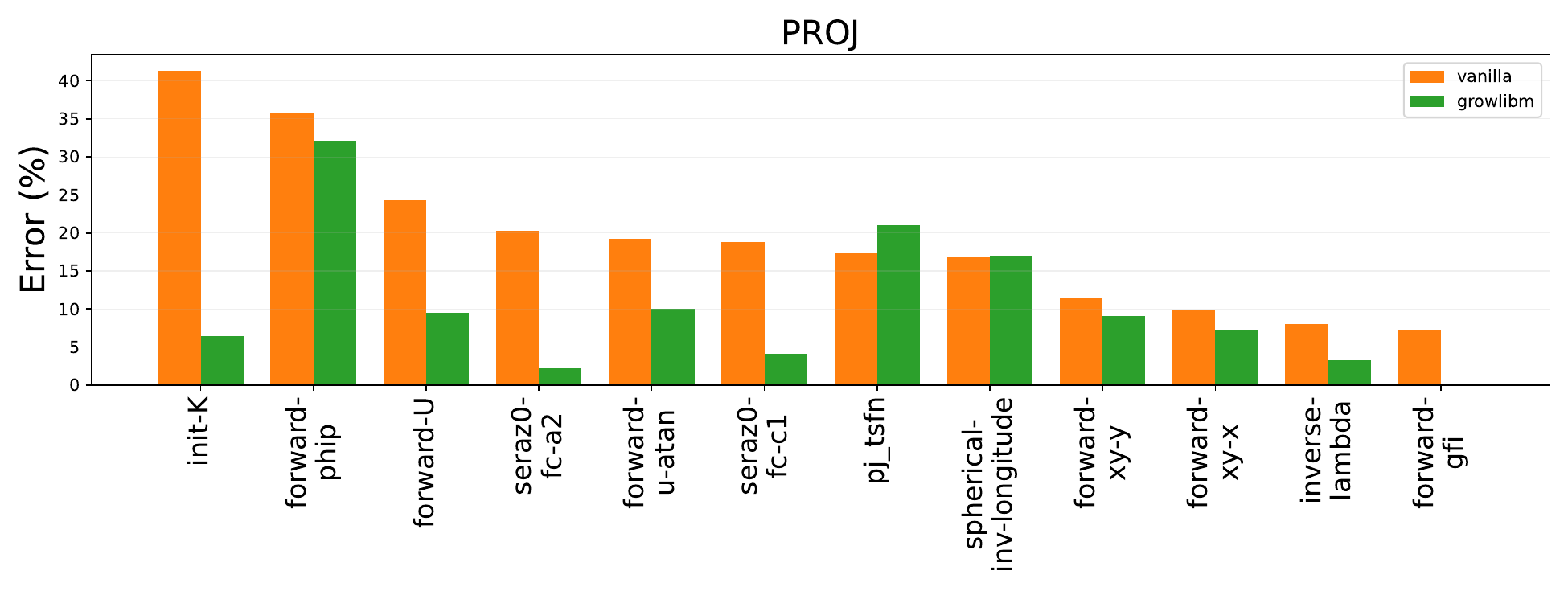}
\includegraphics[width=\columnwidth]{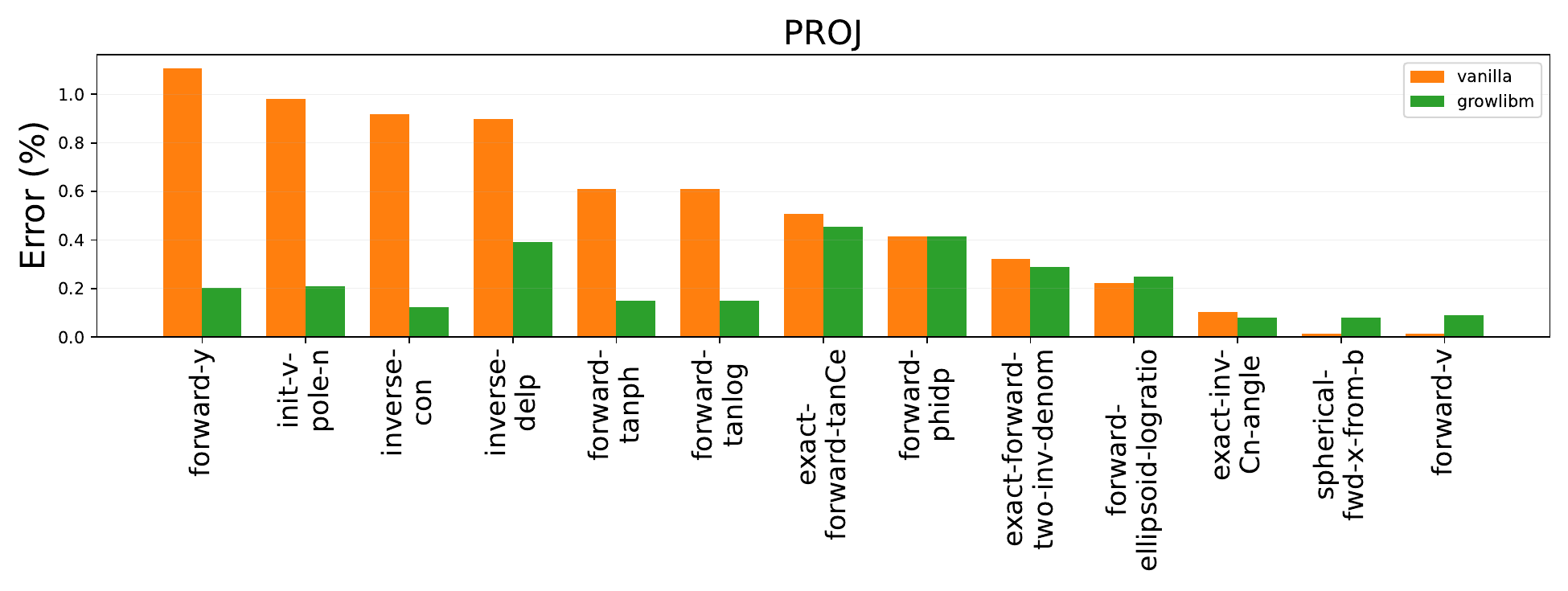} \\
\includegraphics[width=.49\columnwidth]{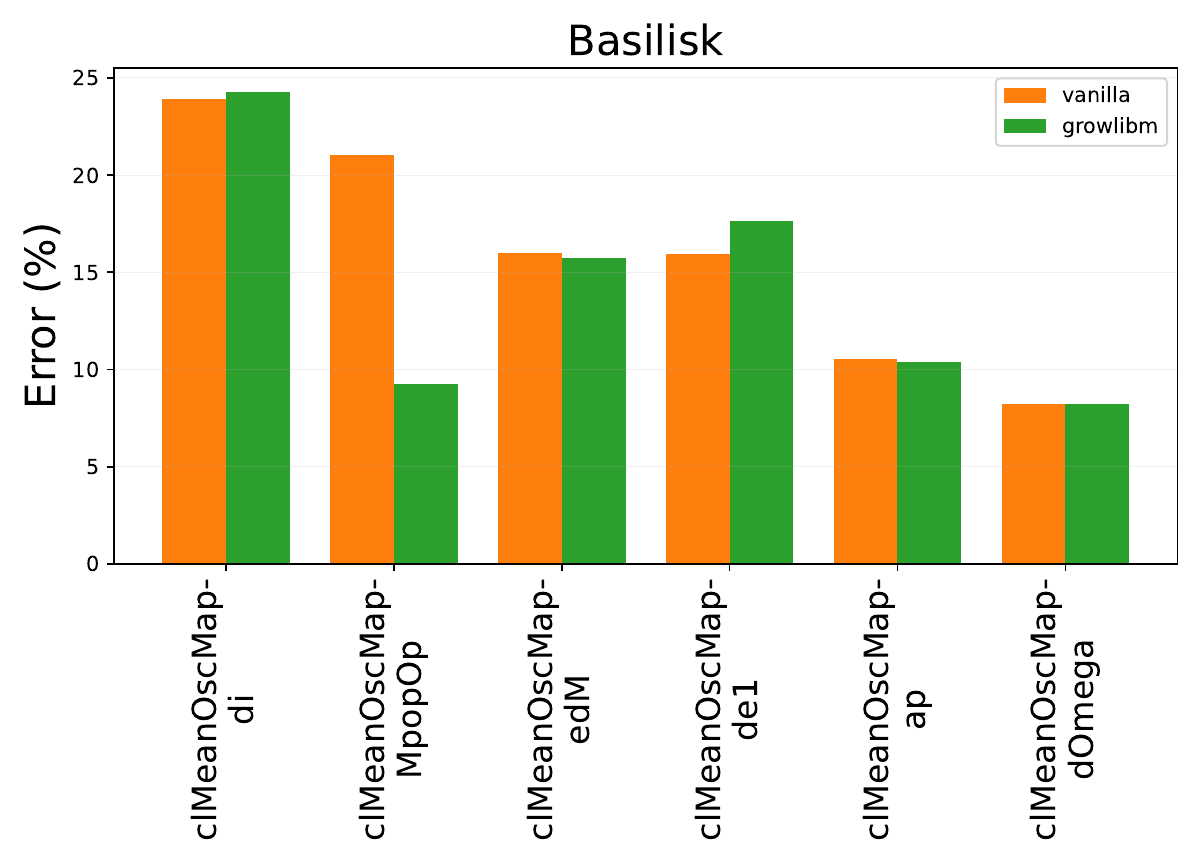}%
\includegraphics[width=.49\columnwidth]{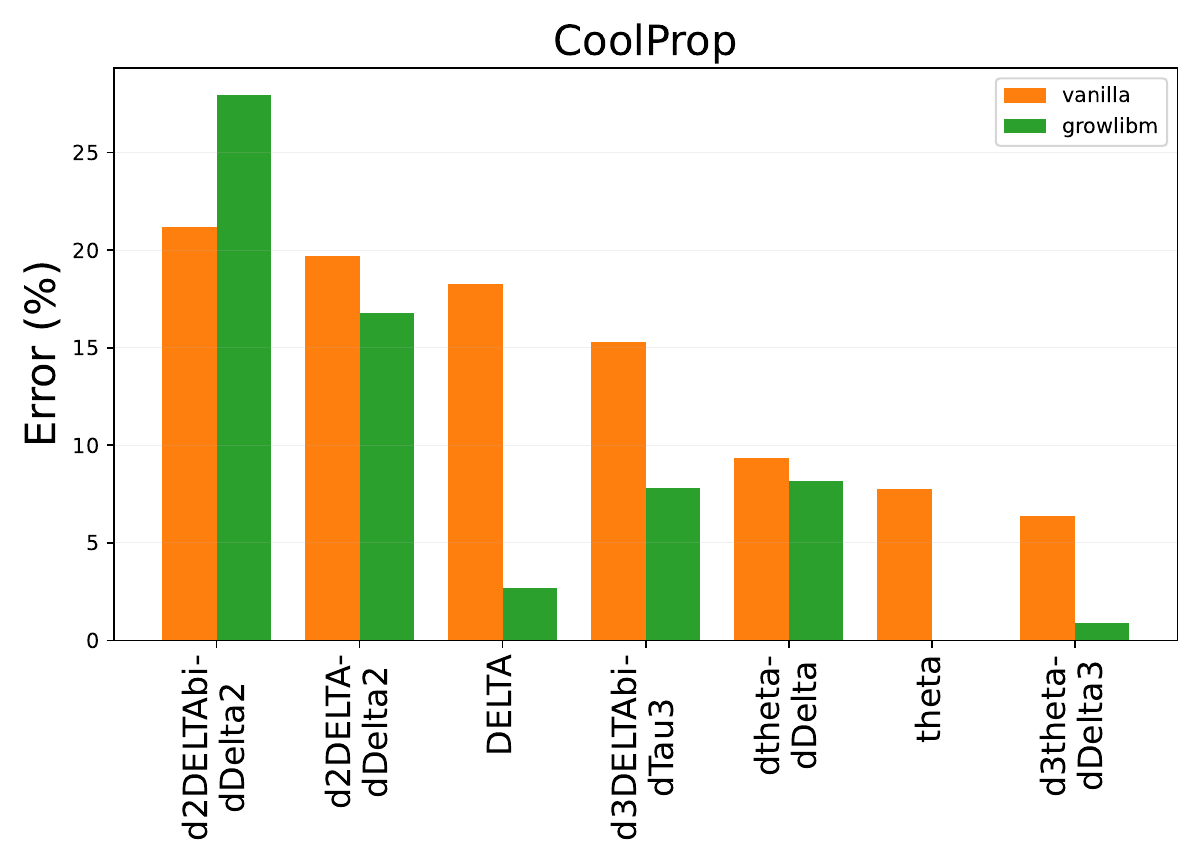}
\caption{
  Lowest error achieved by Herbie
    for kernels in all three target applications
    that use \name-suggested library primitives.
  Error is reported as a percentage of the 64-bit result
    (bits of error divided by 64); lower is better.
  PROJ kernels are split into two charts
    (high-error and low-error) for readability.
  \name-suggested library primitives
    enable dramatic reductions in error
    for most kernels.
}
\label{fig:accuracies}
\end{figure}

Expert implementations of \name-suggested primitives
  should expand the reachable speed/accuracy frontier
  in two ways:
  kernels should be faster at any given accuracy level,
  and the best achievable accuracy should improve.
\Cref{fig:speedup} tests the first claim;
  \Cref{fig:accuracies} tests the second.

\Cref{fig:speedup} compares the speed of each project's kernels
  with and without expert implementations of \name-suggested primitives.
Each curve aggregates all kernels that use
  at least one \name-suggested primitive,
  plotting speedup as a function of accuracy level.
Using the \name-suggested primitives results in
  substantially lower cost at any given level of accuracy.
This is despite the fact that many expert-written implementations
  are slower than naive implementations.
Because these implementations are
  \emph{so} much more accurate than a naive implementation,
  they allow Herbie to reduce the accuracy
  of \emph{other} parts of the kernel,
  recovering the lost performance and then some.
More precisely, on the PROJ benchmarks,
  the \name-enhanced platform definition
  allows Herbie to achieve speedups of
  up to \maxPROJSpeedup for accuracy levels
  above 80\%.
The same happens for CoolProp,
  with speedups of up to \maxCoolPropSpeedup.
For Basilisk,
  where the \name-suggested primitives
  had accurate naive implementations
  but faster expert implementations,
  the speed-up is instead concentrated around
  moderate accuracy levels between 55\%--80\%,
  providing speed-ups of up to \lowerAccuracyBasiliskSpeedup.%
\footnote{
  CoolProp sees 
    slight slow-downs at some accuracies;
    we believe this is mostly noise,
    resulting from the \name-suggested primitives
    causing Herbie to explore its search space differently.
  A user could, of course,
    run Herbie multiple times with different seeds,
    or with and without \name-suggested kernels,
    and pick the best implementation.
}
In short, expert implementations of \name-suggested primitives
  enable substantial speedups at a wide range of accuracies.

The \name-suggested primitives also enable
  lower error than Herbie could achieve on its own.
\Cref{fig:accuracies} demonstrates this
  by plotting the minimum error achieved by Herbie
  on each kernel across all three applications
  where Herbie uses at least one \name-suggested primitive.
On almost all such kernels,
  error is reduced substantially, showing that
  the \name-suggested and expert-implemented primitives
  not only improve speedup at any given level of error
  but also enable lower error than Herbie could achieve on its own.
For a few kernels error \emph{does} increase,
  likely due to how the \name-suggested library primitives
  affect Herbie's internal search process,
  but these increases are small,
  whereas reductions are often dramatic.
For example, in the \texttt{init-K} block from PROJ,
  expert implementations of \name-suggested library primitives
  like \texttt{invgud} and \texttt{log1pmd}
  enable a reduction from roughly 28 bits of error
  (about 44\% error, or 56\% accuracy),
  nearly unusable unless $|\phi|$ and $|\phi_p|$ are close to $1$,
  to roughly 4 bits of error
  (about 6\% error, or 94\% accuracy).
Herbie measures error in bits,
  where each bit of error corresponds to
  one bit of the result being unreliable;
  \Cref{fig:accuracies} plots
  error as a percentage of the 64-bit result,
  i.e., bits of error divided by 64.

\subsection{RQ5: Can Learned Primitives Be Deployed in Compiled Applications?}
\label{subsec:RQ5}

RQ1--RQ4 demonstrate that \name's primitives
  are useful at the kernel level;
  this question asks whether they can also be recognized
  and deployed in compiled application code.

Testing whole-application performance requires
  integrating with the target application build system.
Since doing so with Herbie would be exceptionally challenging,
  we instead developed an LLVM pass
  that rewrites LLVM IR to use \name-selected primitives.
Like Herbie, this pass can apply algebraic identities and rewrites,
  though it is necessarily more limited due to operating
  inside the LLVM compilation pipeline.

\begin{figure}
\begin{tabular}{l | r r r | r}
Projection & \texttt{log1pmd} & \texttt{invgud} & \texttt{hypot} & Total \\
\hline
krovak & 0 & 0 & 0 & 0 \\
omerc & 1  & 1 & 0 & 2 \\
som & 0 & 1 & 4 & 5 \\
somerc & 8 & 10 & 0 & 18 \\
tmerc & 1 & 0 & 0 & 1 \\
\hline
Total & 10 & 12 & 4 & 26 \\
\end{tabular}
\caption{
  Number of replacements performed by the prototype LLVM pass
  for various \name-selected primitives across five PROJ projections:
  Krovak (krovak), Oblique Mercator (omerc),
  Space Oblique Mercator (som), Swiss Oblique Mercator (somerc),
  and Transverse Mercator (tmerc).}
\label{fig:LLVM-replacements}
\end{figure}

The pass runs on each LLVM basic block,
  iterating over each IR instruction
  and uses LLVM's pattern matcher
  to select specific floating-point instructions.
For each instruction,
  we compare the instruction DAG rooted at that instruction
  to a sequence of patterns and, if it matches,
  rewrite it to a call to one of the \name-selected primitives.
The patterns allow for floating-point reassociation and distribution laws;
  for example, the expression $(2x+\pi)/4$ (part of \texttt{invgud})
  is often optimized to $0.5x + \pi/4$ or 
  $\texttt{fma}(x, 0.5, \pi/4)$, where $\pi/4$ is a constant.
Our LLVM pass matches against all of these patterns.
When a match is found, the instruction is replaced with
  a call instruction naming the corresponding \name-suggested library primitive.
We emphasize that this LLVM pass is a prototype,
  far more limited than Herbie itself,
  and incorporated only the three \name-selected primitives
  relevant to PROJ (\texttt{log1pmd}, \texttt{invgud}, and \texttt{hypot}).
It also made no attempts to incorporate a cost model, unlike Herbie.
The results with this LLVM pass are therefore intended
  to demonstrate that \name-selected primitives
  can be recognized and deployed in compiled applications,
  with measurable effects on both speed and accuracy.

\begin{figure}
\includegraphics[width=0.8\columnwidth]{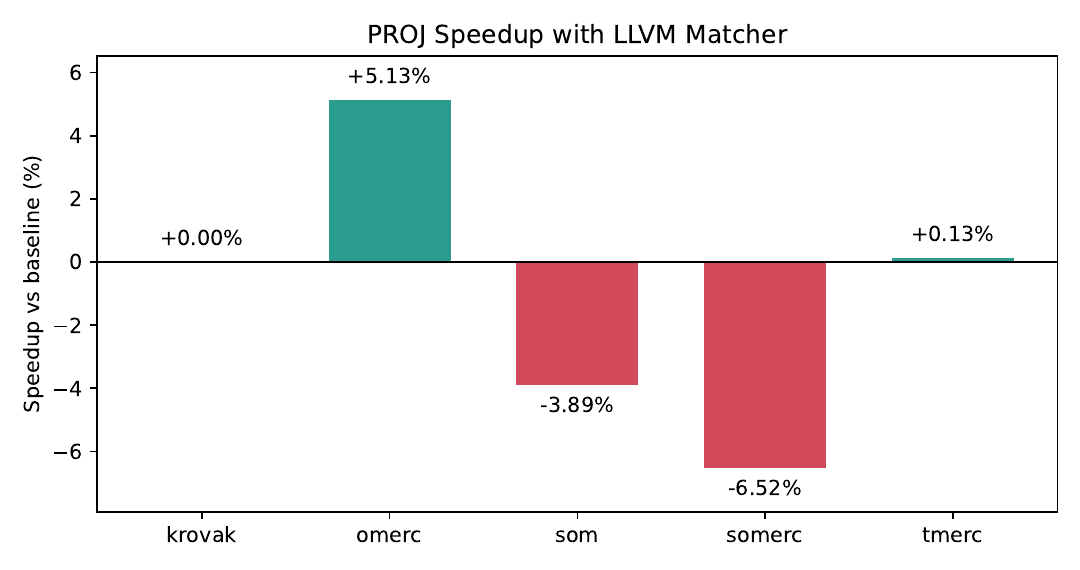}
\caption{Speedup for five PROJ projections,
  compiled using our prototype LLVM pass
  and tested across 10,000,000 points.
  Higher is better.}
\label{fig:LLVM-speedup}
\end{figure}

\Cref{fig:LLVM-replacements} counts how many replacements
  the pass performed across the five projections.
The matcher found 26 use sites,
  fewer than the corresponding Herbie uses in \Cref{fig:eval:function-table}
  because pattern matching in LLVM IR is
  fundamentally more limited than superoptimization.
We built the PROJ application using our new LLVM pass,
  linked it against our library of expert implementations,
  and connected the result to a driver program
  that executes the five cartographic projections
  with 10 million points each.
We then measured the runtime of the resulting program
  on an Apple M3 Pro Macbook Pro with 18 GB of RAM,
  running macOS 15.3.1 and compiling with Apple Clang 17.0.0.

\Cref{fig:LLVM-speedup} shows the results.
The Oblique Mercator (omerc) and Transverse Mercator (tmerc) projections
  see speedups of $5.1\%$ and $0.1\%$ respectively,
  while also benefiting from the higher accuracy
  of the \name-selected primitive implementations.
These are straight wins: faster \emph{and} more accurate.
The Space Oblique Mercator (som) and Swiss Oblique Mercator (somerc),
  which have more replacement sites
  (5 and 18 respectively, mostly \texttt{invgud}),
  see slowdowns of $3.9\%$ and $6.5\%$.
This is because \texttt{invgud}'s expert implementation
  is substantially more accurate
  but slower than the naive code it replaces.
Whether this accuracy-for-speed tradeoff is worthwhile
  is an application-level decision;
  for a cartographic library where geospatial accuracy matters,
  it may well be.
The Krovak (krovak) projection had no replacement sites
  and saw no change.

In short, the LLVM matcher demonstrates
  that \name-selected primitives can be recognized
  and deployed in compiled applications,
  with measurable effects on both speed and accuracy.
Integrating the LLVM pass with a cost model
  could allow it to selectively apply replacements
  only where the tradeoff is favorable.

\subsection{RQ6: How long does \name take to run?}

For \name to be practical,
  the full pipeline must run in reasonable time.

\begin{figure}
\includegraphics[width=\columnwidth]{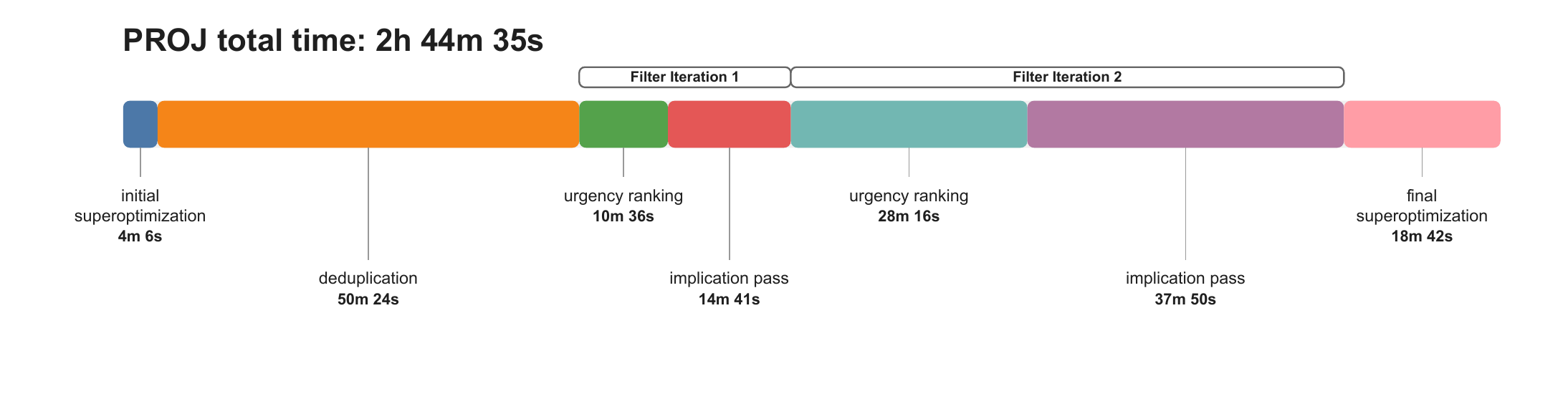}\
\includegraphics[width=\columnwidth]{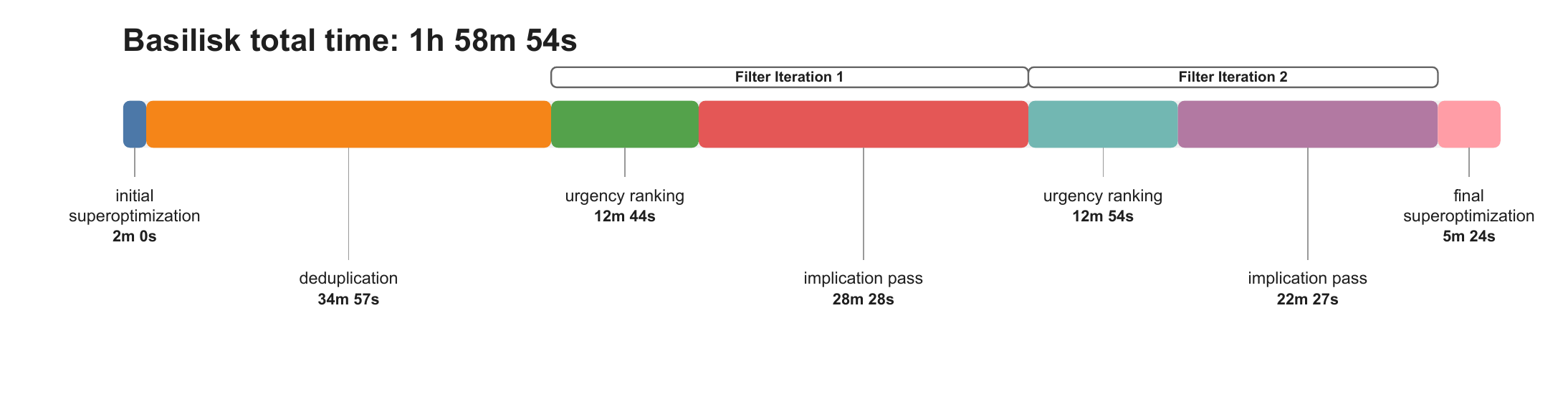}\
\includegraphics[width=\columnwidth]{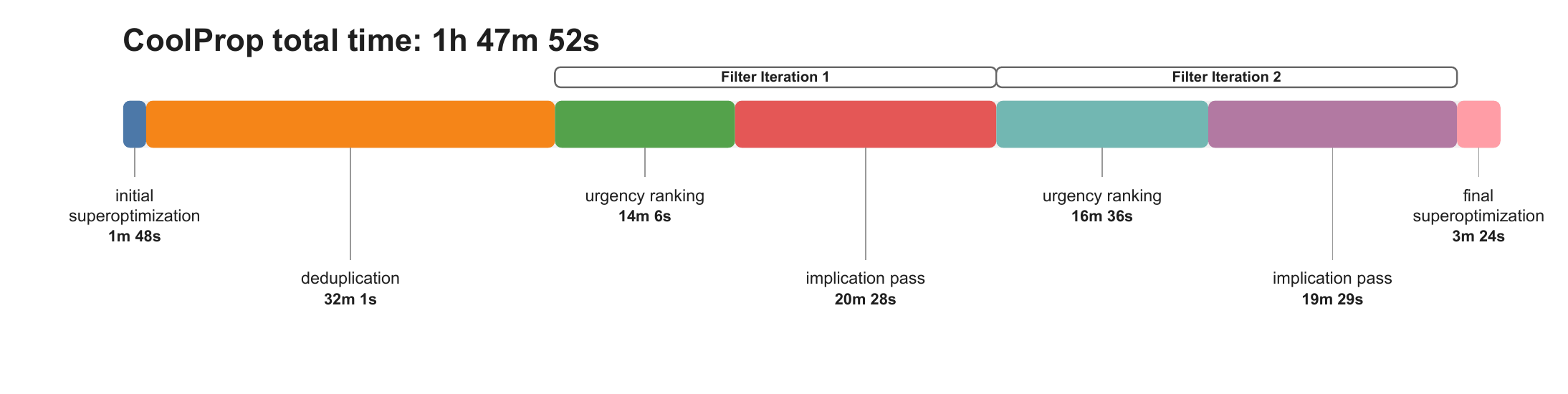}%

\caption{Timeline of \name's stages on the PROJ, Basilisk, and CoolProp applications.}
\label{fig:timeline}
\end{figure}
 
\Cref{fig:timeline} shows \name's runtime
  on the three case studies.
The longest run, on PROJ, takes a little under three hours.
We found this runtime acceptable for the task at hand:
  implementing the selected primitives would surely take longer---%
  in fact, just testing them on billions of inputs took us longer---%
  and would not need to be repeated often.
\Cref{fig:timeline} further subdivides that runtime
  into seven separate steps:
  initial superoptimization with Herbie,
  dumping all generated candidates;
  candidate deduplication, which is the slowest single step;
  two selection stages, each requiring urgency ranking and implication passes;
  and then a final superoptimization using the selected primitives.
This subdivision highlights the importance
  of \name's size, urgency, and frequency heuristics,
  which dramatically reduce the number of superoptimizations needed
  for the two selection stages and thus the time taken by \name.
The two selection stages already take up
  a majority of \name's running time.
We project that a simple greedy selection algorithm,
  as described in \Cref{sec:filter},
  would take several weeks in the same circumstances.
Note that for PROJ, 
  the second selection stage,
  and the final superoptimization,
  are substantially slower than the first selection stage
  and initial superoptimization.
This is due to the optimistic accuracy models used for selected candidates:
  \name uses the Rival interval arithmetic library
  inside these cost models,
  which adds substantial runtime.

\section{Related Work}
\label{sec:relwork}

This section reviews prior work on library learning,
  math library synthesis,
  egraph-based systems, and superoptimizers, and
  clarifies how our work relates to and differs from these lines of research.

\subsection{Library Learning}
Library learning is a technique for
  extracting common structure from a corpus
  of input programs.
Dreamcoder~\cite{dreamcoder}
  uses library learning to generate abstractions that
  can then be used to solve other tasks.
It uses version spaces~\cite{lau} to efficiently
  store the corpus of programs.
Subsequent work on library learning~\cite{babble}
  has used egraph-based anti-unification
  for candidate generation, incorporating equational theories
  to make library learning more robust to syntactic changes.
It also uses an egraph extraction algorithm that relies
  on common subexpression elimination.
Stitch~\cite{stitch} is another library learning tool that
  provides optimality guarantees about the discovered abstractions
  but does not support equational theories.
\name uses a generate and select loop to learn
  abstractions, similar to prior tools~\cite{babble}.

While library learning finds abstractions
  from a set of programs,
  prior work has also focused on other forms 
  of abstraction discovery.
For example, Shapemod~\cite{shapemod} learns
  macros for expressing 3D shapes and
  Szalinski~\cite{szalinski} uses egraph-based
  rewriting to discover maps and folds from
  flattened 3D models.
The principle behind tools like Szalinski is similar
  to \name in that
  they leverage program optimization via rewriting to
  discover new abstractions.

\subsection{Automatic Math Library Generation}
While \name does not generate implementations of the primitives it suggests,
  it complements prior work on automatically constructing such implementations. 
MegaLibm~\cite{megalibm} presents a domain-specific language
  for implementing such primitives.
The key idea in MegaLibm is the separation of
  high-level algorithmic structure from low-level
  floating-point tuning decisions.
Other tools like
  RLIBM~\cite{rlibm, rlibm-all} automatically 
  generate implementations of correctly-rounded math library functions 
  using linear programming.

These systems focus on \textit{how} to
  implement a chosen primitive.
\name addresses the orthogonal and earlier problem of
  \textit{which} primitives to implement to begin with.
This highlights \name's utility in conjunction with 
  automatic implementation tools.

\subsection{Synthesis and Optimization using E-graphs}
The egraph data structure, which was originally
  developed for efficiently representing
  congruence relations in theorem provers~\cite{phd79-egraph},
  has been shown to be effective for program optimization
  via equality saturation~\cite{eqsat}.
The \texttt{egg} library~\cite{egg} offers
  a fast and reusable egraph library that has been
  used in a wide range of systems~\cite{ruler, diospyros, herbie, thesy, babble, dreamcoder,
  3LA, glenside, lakeroad, szalinski, pherbie, isaria, shapecoder, chassis, enumo, spores}.
Even though these systems are from diverse domains,
  they all leverage equality saturation for tasks like
  program synthesis, optimization,
  and equivalence checking.

Egraph-driven rewrite rule synthesis tools
  like Ruler~\cite{ruler} and Enumo~\cite{enumo}
  roughly follow a generate and select loop like \name.
Both Ruler and Enumo target the synthesis of rewrite rules
  whereas \name identifies latent mathematical primitives
  that can expand the frontier of optimal speed-accuracy tradeoffs for
  kernel implementations.
\name relies heavily on Herbie's egraph-based features~\cite{herbie}.
Namely, \name's candidate generation uses Herbie's
  egraph-based term rewriting to search over equivalent terms, 
  and \name's deduplication uses Herbie's 
  egraph-based equivalence checking.





\subsection{Superoptimizing Compilers}
A superoptimizing compiler performs
  a search over a space of programs
  to generate better code~\cite{fraser, superoptimizer}. 
This is in contrast with a normal optimizing compiler 
  which applies a fixed set of transformations.
There has been decades of research on developing superoptimizers.
\citet{bansal} developed a superoptimizer
  for x86 assembly code and relied on exhaustive enumeration
  for exploring the space of programs and canonicalization for 
  pruning the search space.
STOKE~\cite{stoke, stoke-float} is a superoptimizer that relies 
  on stochastic search to effectively search the space of programs.
The Denali superoptimizer~\cite{denali} first
  showed how to use egraph-based rewriting
  to implement a superoptimizer.
Souper~\cite{souper} is a
   middle-end superoptimizer for LLVM.
It uses program synthesis to discover concrete optimizations and
  relies on clever caching of synthesized results for scalability.
Minotaur~\cite{minotaur}
  builds on Souper and uses program synthesis
  to search for low-cost components.
\name's key idea is to repurpose machinery that 
  is common to superoptimizers to 
  discover reusable primitives that can
  unlock better implementations of multiple kernels.
  This, to the best of our knowledge, 
  has not been explored in prior work.
  

\section{Conclusion}

This paper formulates numerical library learning:
  given a workload of floating-point kernels,
  which new primitives are worth expert implementation?
Numerical superoptimizers turn out to already have
  the right machinery for this problem.
Their search enumerates candidates,
  their equivalence reasoning deduplicates and generalizes,
  and their cost models estimate counterfactual utility.
\name puts this observation to work
  by repurposing Herbie as a library learner.

Across PROJ, CoolProp, and Basilisk,
  \name identifies compact primitives
  like \texttt{log1pmd}, \texttt{invgud}, and \texttt{pow1ms}
  that an expert can implement
  using standard numerical techniques.
Herbie reuses these implementations extensively,
  expanding the speed/accuracy frontier
  by up to \maxSpeedupAll at fixed accuracy
  and improving accuracy from \initkStartAccuracy
  to \initkEndAccuracy in the best case.
A prototype LLVM matcher shows that
  learned primitives can also be recognized
  and deployed in compiled code.

Our approach has clear limitations.
\name's ranking heuristics do not fully account for
  implementation difficulty:
  two of the suggested primitives (\texttt{sinquot} and \texttt{cosquot})
  turned out to require prohibitively expensive multi-word arithmetic
  despite ranking well on frequency, urgency, and size.
Incorporating an implementation-difficulty estimate
  into the ranking could address this.
The LLVM deployment prototype lacks a cost model
  and cannot yet make per-site speed/accuracy tradeoffs;
  integrating it more closely with Herbie
  would allow selective deployment
  where the tradeoff is favorable.
And while we believe the library-learning framework
  generalizes beyond numerics,
  our evaluation uses only Herbie
  and only numerical workloads.

These limitations also suggest directions for future work.
The current pipeline requires expert guidance
  to implement the suggested primitives;
  code-generation models could partially automate this step,
  especially for primitives whose implementations
  follow standard patterns like range reduction
  and polynomial approximation.
Testing library learning with other superoptimizers
  is also a natural next step.
Tools like STOKE~\cite{stoke} for x86,
  Souper~\cite{souper} and Minotaur~\cite{minotaur} for LLVM IR,
  and Denali~\cite{denali} for egraph-based rewriting
  all share the core machinery
  (search, equivalence reasoning, cost models)
  that \name repurposes from Herbie.
Combining workload-driven primitive discovery
  with automated implementation via coding models
  could eventually close the loop entirely:
  a superoptimizer that not only identifies
  which primitives would help,
  but also generates and validates their implementations.

At its core, this paper shows that
  the question of which primitives to build
  can be answered by the same machinery
  that optimizes code against existing primitives.
In the three workloads we studied,
  that machinery identified primitives
  that were compact, reusable, implementable,
  and genuinely useful.

\appendix
\section{Data Availability Statement }
\label{sec:availability}

We will submit \name’s source code
  and the scripts used to produce the results in \autoref{sec:eval}
  as part of the OOPSLA artifact evaluation process,
  together with documentation describing how to run them.
All benchmarks used in our evaluation are open source.
The code, scripts, and benchmark set will be made publicly available.

\bibliographystyle{ACM-Reference-Format}
\bibliography{../references.bib}


\begin{thebibliography}{60}


\ifx \showCODEN    \undefined \def \showCODEN     #1{\unskip}     \fi
\ifx \showDOI      \undefined \def \showDOI       #1{#1}\fi
\ifx \showISBNx    \undefined \def \showISBNx     #1{\unskip}     \fi
\ifx \showISBNxiii \undefined \def \showISBNxiii  #1{\unskip}     \fi
\ifx \showISSN     \undefined \def \showISSN      #1{\unskip}     \fi
\ifx \showLCCN     \undefined \def \showLCCN      #1{\unskip}     \fi
\ifx \shownote     \undefined \def \shownote      #1{#1}          \fi
\ifx \showarticletitle \undefined \def \showarticletitle #1{#1}   \fi
\ifx \showURL      \undefined \def \showURL       {\relax}        \fi
\providecommand\bibfield[2]{#2}
\providecommand\bibinfo[2]{#2}
\providecommand\natexlab[1]{#1}
\providecommand\showeprint[2][]{arXiv:#2}

\bibitem[Bansal and Aiken(2006)]%
        {bansal}
\bibfield{author}{\bibinfo{person}{Sorav Bansal} {and} \bibinfo{person}{Alex
  Aiken}.} \bibinfo{year}{2006}\natexlab{}.
\newblock \showarticletitle{Automatic generation of peephole superoptimizers}.
  In \bibinfo{booktitle}{\emph{Proceedings of the 12th International Conference
  on Architectural Support for Programming Languages and Operating Systems}}
  (San Jose, California, USA) \emph{(\bibinfo{series}{ASPLOS XII})}.
  \bibinfo{publisher}{Association for Computing Machinery},
  \bibinfo{address}{New York, NY, USA}, \bibinfo{pages}{394–403}.
\newblock
\showISBNx{1595934510}
\urldef\tempurl%
\url{https://doi.org/10.1145/1168857.1168906}
\showDOI{\tempurl}


\bibitem[Bell et~al\mbox{.}(2014)]%
        {coolprop}
\bibfield{author}{\bibinfo{person}{Ian~H. Bell}, \bibinfo{person}{Jorrit
  Wronski}, \bibinfo{person}{Sylvain Quoilin}, {and} \bibinfo{person}{Vincent
  Lemort}.} \bibinfo{year}{2014}\natexlab{}.
\newblock \showarticletitle{Pure and Pseudo-pure Fluid Thermophysical Property
  Evaluation and the Open-Source Thermophysical Property Library {CoolProp}}.
\newblock \bibinfo{journal}{\emph{Industrial \& Engineering Chemistry
  Research}} \bibinfo{volume}{53}, \bibinfo{number}{6} (\bibinfo{year}{2014}),
  \bibinfo{pages}{2498--2508}.
\newblock
\urldef\tempurl%
\url{https://doi.org/10.1021/ie4033999}
\showDOI{\tempurl}


\bibitem[Ben~Khalifa et~al\mbox{.}(2020)]%
        {pop}
\bibfield{author}{\bibinfo{person}{Dorra Ben~Khalifa},
  \bibinfo{person}{Matthieu Martel}, {and} \bibinfo{person}{Assale Adj\'{e}}.}
  \bibinfo{year}{2020}\natexlab{}.
\newblock \showarticletitle{{POP}: A Tuning Assistant for Mixed-Precision
  Floating-Point Computations}. In \bibinfo{booktitle}{\emph{Formal Techniques
  for Safety-Critical Systems (FTSCS 2019)}}
  \emph{(\bibinfo{series}{Communications in Computer and Information Science},
  Vol.~\bibinfo{volume}{1165})}. \bibinfo{publisher}{Springer},
  \bibinfo{pages}{77--94}.
\newblock
\urldef\tempurl%
\url{https://doi.org/10.1007/978-3-030-46902-3_5}
\showDOI{\tempurl}


\bibitem[Bowers et~al\mbox{.}(2023)]%
        {stitch}
\bibfield{author}{\bibinfo{person}{Matthew Bowers}, \bibinfo{person}{Theo~X.
  Olausson}, \bibinfo{person}{Lionel Wong}, \bibinfo{person}{Gabriel Grand},
  \bibinfo{person}{Joshua~B. Tenenbaum}, \bibinfo{person}{Kevin Ellis}, {and}
  \bibinfo{person}{Armando Solar-Lezama}.} \bibinfo{year}{2023}\natexlab{}.
\newblock \showarticletitle{Top-Down Synthesis for Library Learning}.
\newblock \bibinfo{journal}{\emph{Proc. ACM Program. Lang.}}
  \bibinfo{volume}{7}, \bibinfo{number}{POPL}, Article \bibinfo{articleno}{41}
  (\bibinfo{date}{Jan.} \bibinfo{year}{2023}), \bibinfo{numpages}{32}~pages.
\newblock
\urldef\tempurl%
\url{https://doi.org/10.1145/3571234}
\showDOI{\tempurl}


\bibitem[Briggs et~al\mbox{.}(2024)]%
        {megalibm}
\bibfield{author}{\bibinfo{person}{Ian Briggs}, \bibinfo{person}{Yash Lad},
  {and} \bibinfo{person}{Pavel Panchekha}.} \bibinfo{year}{2024}\natexlab{}.
\newblock \showarticletitle{Implementation and Synthesis of Math Library
  Functions}.
\newblock \bibinfo{journal}{\emph{Proc. ACM Program. Lang.}}
  \bibinfo{volume}{8}, \bibinfo{number}{POPL}, Article \bibinfo{articleno}{32}
  (\bibinfo{date}{jan} \bibinfo{year}{2024}), \bibinfo{numpages}{28}~pages.
\newblock
\urldef\tempurl%
\url{https://doi.org/10.1145/3632874}
\showDOI{\tempurl}


\bibitem[Cao et~al\mbox{.}(2023)]%
        {babble}
\bibfield{author}{\bibinfo{person}{David Cao}, \bibinfo{person}{Rose Kunkel},
  \bibinfo{person}{Chandrakana Nandi}, \bibinfo{person}{Max Willsey},
  \bibinfo{person}{Zachary Tatlock}, {and} \bibinfo{person}{Nadia
  Polikarpova}.} \bibinfo{year}{2023}\natexlab{}.
\newblock \showarticletitle{babble: Learning Better Abstractions with E-Graphs
  and Anti-unification}.
\newblock \bibinfo{journal}{\emph{Proc. ACM Program. Lang.}}
  \bibinfo{volume}{7}, \bibinfo{number}{POPL}, Article \bibinfo{articleno}{14}
  (\bibinfo{date}{Jan.} \bibinfo{year}{2023}), \bibinfo{numpages}{29}~pages.
\newblock
\urldef\tempurl%
\url{https://doi.org/10.1145/3571207}
\showDOI{\tempurl}


\bibitem[Chevillard et~al\mbox{.}(2010)]%
        {sollya}
\bibfield{author}{\bibinfo{person}{Sylvain Chevillard}, \bibinfo{person}{Mioara
  Jolde\c{s}}, {and} \bibinfo{person}{Christoph Lauter}.}
  \bibinfo{year}{2010}\natexlab{}.
\newblock \showarticletitle{Sollya: An Environment for the Development of
  Numerical Codes}. In \bibinfo{booktitle}{\emph{Proceedings of the Third
  International Congress on Mathematical Software}}
  \emph{(\bibinfo{series}{ICMS'10})}. \bibinfo{publisher}{Springer-Verlag},
  \bibinfo{pages}{28--31}.
\newblock
\showISBNx{3642155812}


\bibitem[Chiang et~al\mbox{.}(2017)]%
        {fptuner}
\bibfield{author}{\bibinfo{person}{Wei-Fan Chiang}, \bibinfo{person}{Mark
  Baranowski}, \bibinfo{person}{Ian Briggs}, \bibinfo{person}{Alexey Solovyev},
  \bibinfo{person}{Ganesh Gopalakrishnan}, {and} \bibinfo{person}{Zvonimir
  Rakamari\'c}.} \bibinfo{year}{2017}\natexlab{}.
\newblock \showarticletitle{Rigorous Floating-point Mixed-precision Tuning}
  \emph{(\bibinfo{series}{POPL})}. \bibinfo{pages}{300--315}.
\newblock
\showISBNx{978-1-4503-4660-3}
\urldef\tempurl%
\url{https://doi.org/10.1145/3009837.3009846}
\showDOI{\tempurl}


\bibitem[Damouche and Martel(2017)]%
        {salsa}
\bibfield{author}{\bibinfo{person}{Nasrine Damouche} {and}
  \bibinfo{person}{Matthieu Martel}.} \bibinfo{year}{2017}\natexlab{}.
\newblock \showarticletitle{Salsa: An automatic tool to improve the numerical
  accuracy of programs} \emph{(\bibinfo{series}{AFM})}.
\newblock


\bibitem[Damouche et~al\mbox{.}(2017)]%
        {salsa-sttt}
\bibfield{author}{\bibinfo{person}{Nasrine Damouche}, \bibinfo{person}{Matthieu
  Martel}, {and} \bibinfo{person}{Alexandre Chapoutot}.}
  \bibinfo{year}{2017}\natexlab{}.
\newblock \showarticletitle{Improving the Numerical Accuracy of Programs by
  Automatic Transformation}.
\newblock \bibinfo{journal}{\emph{International Journal on Software Tools for
  Technology Transfer}} \bibinfo{volume}{19}, \bibinfo{number}{4}
  (\bibinfo{year}{2017}), \bibinfo{pages}{427--448}.
\newblock
\urldef\tempurl%
\url{https://doi.org/10.1007/s10009-016-0435-0}
\showDOI{\tempurl}


\bibitem[Damouche et~al\mbox{.}(2016)]%
        {fpbench}
\bibfield{author}{\bibinfo{person}{Nasrine Damouche}, \bibinfo{person}{Matthieu
  Martel}, \bibinfo{person}{Pavel Panchekha}, \bibinfo{person}{Jason Qiu},
  \bibinfo{person}{Alex Sanchez-Stern}, {and} \bibinfo{person}{Zachary
  Tatlock}.} \bibinfo{year}{2016}\natexlab{}.
\newblock \showarticletitle{Toward a Standard Benchmark Format and Suite for
  Floating-Point Analysis}.
\newblock  (\bibinfo{date}{July} \bibinfo{year}{2016}).
\newblock


\bibitem[Darulova and Kuncak(2014)]%
        {rosa}
\bibfield{author}{\bibinfo{person}{Eva Darulova} {and} \bibinfo{person}{Viktor
  Kuncak}.} \bibinfo{year}{2014}\natexlab{}.
\newblock \showarticletitle{Sound Compilation of Reals}
  \emph{(\bibinfo{series}{POPL})}. \bibinfo{numpages}{14}~pages.
\newblock
\showISBNx{978-1-4503-2544-8}
\urldef\tempurl%
\url{http://doi.acm.org/10.1145/2535838.2535874}
\showURL{%
\tempurl}


\bibitem[Das et~al\mbox{.}(2020)]%
        {satire}
\bibfield{author}{\bibinfo{person}{Arnab Das}, \bibinfo{person}{Ian Briggs},
  \bibinfo{person}{Ganesh Gopalakrishnan}, \bibinfo{person}{Sriram
  Krishnamoorthy}, {and} \bibinfo{person}{Pavel Panchekha}.}
  \bibinfo{year}{2020}\natexlab{}.
\newblock \showarticletitle{Scalable yet Rigorous Floating-Point Error
  Analysis}. In \bibinfo{booktitle}{\emph{2020 SC20: International Conference
  for High Performance Computing, Networking, Storage and Analysis (SC)}}.
  \bibinfo{publisher}{IEEE Computer Society}, \bibinfo{address}{Los Alamitos,
  CA, USA}, \bibinfo{pages}{1--14}.
\newblock
\urldef\tempurl%
\url{https://doi.org/10.1109/SC41405.2020.00055}
\showDOI{\tempurl}


\bibitem[Ellis et~al\mbox{.}(2021)]%
        {dreamcoder}
\bibfield{author}{\bibinfo{person}{Kevin Ellis}, \bibinfo{person}{Catherine
  Wong}, \bibinfo{person}{Maxwell Nye}, \bibinfo{person}{Mathias
  Sabl\'{e}-Meyer}, \bibinfo{person}{Lucas Morales}, \bibinfo{person}{Luke
  Hewitt}, \bibinfo{person}{Luc Cary}, \bibinfo{person}{Armando Solar-Lezama},
  {and} \bibinfo{person}{Joshua~B. Tenenbaum}.}
  \bibinfo{year}{2021}\natexlab{}.
\newblock \showarticletitle{DreamCoder: bootstrapping inductive program
  synthesis with wake-sleep library learning}. In
  \bibinfo{booktitle}{\emph{Proceedings of the 42nd ACM SIGPLAN International
  Conference on Programming Language Design and Implementation}} (Virtual,
  Canada) \emph{(\bibinfo{series}{PLDI 2021})}. \bibinfo{publisher}{Association
  for Computing Machinery}, \bibinfo{address}{New York, NY, USA},
  \bibinfo{pages}{835–850}.
\newblock
\showISBNx{9781450383912}
\urldef\tempurl%
\url{https://doi.org/10.1145/3453483.3454080}
\showDOI{\tempurl}


\bibitem[Flatt et~al\mbox{.}(2022)]%
        {egg-proofs}
\bibfield{author}{\bibinfo{person}{Oliver Flatt}, \bibinfo{person}{Samuel
  Coward}, \bibinfo{person}{Max Willsey}, \bibinfo{person}{Zachary Tatlock},
  {and} \bibinfo{person}{Pavel Panchekha}.} \bibinfo{year}{2022}\natexlab{}.
\newblock \showarticletitle{Small Proofs from Congruence Closure}. In
  \bibinfo{booktitle}{\emph{Proceedings of the 22nd Conference on Formal
  Methods in Computer-Aided Design (FMCAD)}}. \bibinfo{publisher}{TU Wien
  Academic Press}, \bibinfo{pages}{75--83}.
\newblock
\urldef\tempurl%
\url{https://doi.org/10.34727/2022/isbn.978-3-85448-053-2_13}
\showDOI{\tempurl}


\bibitem[Flatt and Panchekha(2023)]%
        {movability}
\bibfield{author}{\bibinfo{person}{Oliver Flatt} {and} \bibinfo{person}{Pavel
  Panchekha}.} \bibinfo{year}{2023}\natexlab{}.
\newblock \showarticletitle{Making Interval Arithmetic Robust to Overflow}. In
  \bibinfo{booktitle}{\emph{2023 IEEE 30th Symposium on Computer Arithmetic
  (ARITH)}}. \bibinfo{pages}{44--47}.
\newblock
\urldef\tempurl%
\url{https://doi.org/10.1109/ARITH58626.2023.00022}
\showDOI{\tempurl}


\bibitem[Flatt and Zhang(2023)]%
        {egglog-in-practice}
\bibfield{author}{\bibinfo{person}{Oliver Flatt} {and} \bibinfo{person}{Yihong
  Zhang}.} \bibinfo{year}{2023}\natexlab{}.
\newblock \showarticletitle{egglog In Practice: Automatically Improving
  Floating-point Error}. In \bibinfo{booktitle}{\emph{EGRAPHS 2023: E-Graph
  Research, Applications, Practices, and Human-factors Symposium}}.
\newblock
\urldef\tempurl%
\url{https://effect.systems/doc/egraphs-2023-egglog/paper.pdf}
\showURL{%
\tempurl}


\bibitem[{FPBench Project}(2021)]%
        {fpcore}
\bibfield{author}{\bibinfo{person}{{FPBench Project}}.}
  \bibinfo{year}{2021}\natexlab{}.
\newblock \bibinfo{title}{{FPCore} 2.0: The Standard Floating-Point Benchmark
  Format}.
\newblock
\newblock
\urldef\tempurl%
\url{https://fpbench.org/spec/fpcore-2.0.html}
\showURL{%
\tempurl}
\newblock
\shownote{Accessed March 2026}.


\bibitem[Fraser(1979)]%
        {fraser}
\bibfield{author}{\bibinfo{person}{Christopher~W. Fraser}.}
  \bibinfo{year}{1979}\natexlab{}.
\newblock \showarticletitle{A compact, machine-independent peephole optimizer}.
  In \bibinfo{booktitle}{\emph{Proceedings of the 6th ACM SIGACT-SIGPLAN
  Symposium on Principles of Programming Languages}} (San Antonio, Texas)
  \emph{(\bibinfo{series}{POPL '79})}. \bibinfo{publisher}{Association for
  Computing Machinery}, \bibinfo{address}{New York, NY, USA},
  \bibinfo{pages}{1–6}.
\newblock
\showISBNx{9781450373579}
\urldef\tempurl%
\url{https://doi.org/10.1145/567752.567753}
\showDOI{\tempurl}


\bibitem[Huang et~al\mbox{.}(2024)]%
        {3LA}
\bibfield{author}{\bibinfo{person}{Bo-Yuan Huang}, \bibinfo{person}{Steven
  Lyubomirsky}, \bibinfo{person}{Yi Li}, \bibinfo{person}{Mike He},
  \bibinfo{person}{Gus~Henry Smith}, \bibinfo{person}{Thierry Tambe},
  \bibinfo{person}{Akash Gaonkar}, \bibinfo{person}{Vishal Canumalla},
  \bibinfo{person}{Andrew Cheung}, \bibinfo{person}{Gu-Yeon Wei},
  \bibinfo{person}{Aarti Gupta}, \bibinfo{person}{Zachary Tatlock}, {and}
  \bibinfo{person}{Sharad Malik}.} \bibinfo{year}{2024}\natexlab{}.
\newblock \showarticletitle{Application-level Validation of Accelerator Designs
  Using a Formal Software/Hardware Interface}.
\newblock \bibinfo{journal}{\emph{ACM Trans. Des. Autom. Electron. Syst.}}
  \bibinfo{volume}{29}, \bibinfo{number}{2}, Article \bibinfo{articleno}{35}
  (\bibinfo{date}{Feb.} \bibinfo{year}{2024}), \bibinfo{numpages}{25}~pages.
\newblock
\showISSN{1084-4309}
\urldef\tempurl%
\url{https://doi.org/10.1145/3639051}
\showDOI{\tempurl}


\bibitem[Ioualalen and Martel(2012)]%
        {sardana}
\bibfield{author}{\bibinfo{person}{Arnault Ioualalen} {and}
  \bibinfo{person}{Matthieu Martel}.} \bibinfo{year}{2012}\natexlab{}.
\newblock \showarticletitle{Sardana: An Automatic Tool for Numerical Accuracy
  Optimization}. In \bibinfo{booktitle}{\emph{15th GAMM-IMACS International
  Symposium on Scientific Computing, Computer Arithmetic and Validated Numerics
  (SCAN)}}.
\newblock


\bibitem[Izycheva and Darulova(2017)]%
        {daisy}
\bibfield{author}{\bibinfo{person}{Anastasiia Izycheva} {and}
  \bibinfo{person}{Eva Darulova}.} \bibinfo{year}{2017}\natexlab{}.
\newblock \showarticletitle{On sound relative error bounds for floating-point
  arithmetic} \emph{(\bibinfo{series}{{FMCAD}})}. \bibinfo{pages}{15--22}.
\newblock
\urldef\tempurl%
\url{https://doi.org/10.23919/FMCAD.2017.8102236}
\showDOI{\tempurl}


\bibitem[Jones et~al\mbox{.}(2021)]%
        {shapemod}
\bibfield{author}{\bibinfo{person}{R.~Kenny Jones}, \bibinfo{person}{David
  Charatan}, \bibinfo{person}{Paul Guerrero}, \bibinfo{person}{Niloy~J. Mitra},
  {and} \bibinfo{person}{Daniel Ritchie}.} \bibinfo{year}{2021}\natexlab{}.
\newblock \showarticletitle{ShapeMOD: macro operation discovery for 3D shape
  programs}.
\newblock \bibinfo{journal}{\emph{ACM Trans. Graph.}} \bibinfo{volume}{40},
  \bibinfo{number}{4}, Article \bibinfo{articleno}{153} (\bibinfo{date}{July}
  \bibinfo{year}{2021}), \bibinfo{numpages}{16}~pages.
\newblock
\showISSN{0730-0301}
\urldef\tempurl%
\url{https://doi.org/10.1145/3450626.3459821}
\showDOI{\tempurl}


\bibitem[Jones et~al\mbox{.}(2023)]%
        {shapecoder}
\bibfield{author}{\bibinfo{person}{R.~Kenny Jones}, \bibinfo{person}{Paul
  Guerrero}, \bibinfo{person}{Niloy~J. Mitra}, {and} \bibinfo{person}{Daniel
  Ritchie}.} \bibinfo{year}{2023}\natexlab{}.
\newblock \showarticletitle{ShapeCoder: Discovering Abstractions for Visual
  Programs from Unstructured Primitives}.
\newblock \bibinfo{journal}{\emph{ACM Trans. Graph.}} \bibinfo{volume}{42},
  \bibinfo{number}{4}, Article \bibinfo{articleno}{49} (\bibinfo{date}{July}
  \bibinfo{year}{2023}), \bibinfo{numpages}{17}~pages.
\newblock
\showISSN{0730-0301}
\urldef\tempurl%
\url{https://doi.org/10.1145/3592416}
\showDOI{\tempurl}


\bibitem[Joshi et~al\mbox{.}(2002)]%
        {denali}
\bibfield{author}{\bibinfo{person}{Rajeev Joshi}, \bibinfo{person}{Greg
  Nelson}, {and} \bibinfo{person}{Keith Randall}.}
  \bibinfo{year}{2002}\natexlab{}.
\newblock \showarticletitle{Denali: a goal-directed superoptimizer}. In
  \bibinfo{booktitle}{\emph{Proceedings of the ACM SIGPLAN 2002 Conference on
  Programming Language Design and Implementation}} (Berlin, Germany)
  \emph{(\bibinfo{series}{PLDI '02})}. \bibinfo{publisher}{Association for
  Computing Machinery}, \bibinfo{address}{New York, NY, USA},
  \bibinfo{pages}{304–314}.
\newblock
\showISBNx{1581134630}
\urldef\tempurl%
\url{https://doi.org/10.1145/512529.512566}
\showDOI{\tempurl}


\bibitem[Kenneally et~al\mbox{.}(2020)]%
        {basilisk}
\bibfield{author}{\bibinfo{person}{Patrick~W. Kenneally},
  \bibinfo{person}{Scott Piggott}, {and} \bibinfo{person}{Hanspeter Schaub}.}
  \bibinfo{year}{2020}\natexlab{}.
\newblock \showarticletitle{Basilisk: A Flexible, Scalable and Modular
  Astrodynamics Simulation Framework}.
\newblock \bibinfo{journal}{\emph{Journal of Aerospace Information Systems}}
  \bibinfo{volume}{17}, \bibinfo{number}{9} (\bibinfo{year}{2020}).
\newblock
\urldef\tempurl%
\url{https://doi.org/10.2514/1.I010762}
\showDOI{\tempurl}


\bibitem[Kulkarni and Panchekha(2025)]%
        {explanifloat}
\bibfield{author}{\bibinfo{person}{Bhargav Kulkarni} {and}
  \bibinfo{person}{Pavel Panchekha}.} \bibinfo{year}{2025}\natexlab{}.
\newblock \showarticletitle{Mixing Condition Numbers and Oracles for Accurate
  Floating-point Debugging}. In \bibinfo{booktitle}{\emph{2025 IEEE 32nd
  Symposium on Computer Arithmetic (ARITH)}}.
\newblock


\bibitem[Lau et~al\mbox{.}(2000)]%
        {lau}
\bibfield{author}{\bibinfo{person}{Tessa~A. Lau}, \bibinfo{person}{Pedro
  Domingos}, {and} \bibinfo{person}{Daniel~S. Weld}.}
  \bibinfo{year}{2000}\natexlab{}.
\newblock \showarticletitle{Version Space Algebra and its Application to
  Programming by Demonstration}. In \bibinfo{booktitle}{\emph{Proceedings of
  the Seventeenth International Conference on Machine Learning}}
  \emph{(\bibinfo{series}{ICML '00})}. \bibinfo{publisher}{Morgan Kaufmann
  Publishers Inc.}, \bibinfo{address}{San Francisco, CA, USA},
  \bibinfo{pages}{527–534}.
\newblock
\showISBNx{1558607072}


\bibitem[Lim and Nagarakatte(2021)]%
        {rlibm}
\bibfield{author}{\bibinfo{person}{Jay~P. Lim} {and} \bibinfo{person}{Santosh
  Nagarakatte}.} \bibinfo{year}{2021}\natexlab{}.
\newblock \showarticletitle{High performance correctly rounded math libraries
  for 32-bit floating point representations}. In
  \bibinfo{booktitle}{\emph{Proceedings of the 42nd ACM SIGPLAN International
  Conference on Programming Language Design and Implementation}} (Virtual,
  Canada) \emph{(\bibinfo{series}{PLDI 2021})}. \bibinfo{publisher}{Association
  for Computing Machinery}, \bibinfo{address}{New York, NY, USA},
  \bibinfo{pages}{359–374}.
\newblock
\showISBNx{9781450383912}
\urldef\tempurl%
\url{https://doi.org/10.1145/3453483.3454049}
\showDOI{\tempurl}


\bibitem[Liu et~al\mbox{.}(2024)]%
        {minotaur}
\bibfield{author}{\bibinfo{person}{Zhengyang Liu}, \bibinfo{person}{Stefan
  Mada}, {and} \bibinfo{person}{John Regehr}.} \bibinfo{year}{2024}\natexlab{}.
\newblock \showarticletitle{Minotaur: A SIMD-Oriented Synthesizing
  Superoptimizer}.
\newblock \bibinfo{journal}{\emph{Proc. ACM Program. Lang.}}
  \bibinfo{volume}{8}, \bibinfo{number}{OOPSLA2}, Article
  \bibinfo{articleno}{326} (\bibinfo{date}{Oct.} \bibinfo{year}{2024}),
  \bibinfo{numpages}{25}~pages.
\newblock
\urldef\tempurl%
\url{https://doi.org/10.1145/3689766}
\showDOI{\tempurl}


\bibitem[Massalin(1987)]%
        {superoptimizer}
\bibfield{author}{\bibinfo{person}{Henry Massalin}.}
  \bibinfo{year}{1987}\natexlab{}.
\newblock \showarticletitle{Superoptimizer: a look at the smallest program}. In
  \bibinfo{booktitle}{\emph{Proceedings of the Second International Conference
  on Architectual Support for Programming Languages and Operating Systems}}
  (Palo Alto, California, USA) \emph{(\bibinfo{series}{ASPLOS II})}.
  \bibinfo{publisher}{Association for Computing Machinery},
  \bibinfo{address}{New York, NY, USA}, \bibinfo{pages}{122–126}.
\newblock
\showISBNx{0818608056}
\urldef\tempurl%
\url{https://doi.org/10.1145/36206.36194}
\showDOI{\tempurl}


\bibitem[Nandi et~al\mbox{.}(2020)]%
        {szalinski}
\bibfield{author}{\bibinfo{person}{Chandrakana Nandi}, \bibinfo{person}{Max
  Willsey}, \bibinfo{person}{Adam Anderson}, \bibinfo{person}{James~R. Wilcox},
  \bibinfo{person}{Eva Darulova}, \bibinfo{person}{Dan Grossman}, {and}
  \bibinfo{person}{Zachary Tatlock}.} \bibinfo{year}{2020}\natexlab{}.
\newblock \showarticletitle{Synthesizing structured CAD models with equality
  saturation and inverse transformations}. In
  \bibinfo{booktitle}{\emph{Proceedings of the 41st ACM SIGPLAN Conference on
  Programming Language Design and Implementation}} (London, UK)
  \emph{(\bibinfo{series}{PLDI 2020})}. \bibinfo{publisher}{Association for
  Computing Machinery}, \bibinfo{address}{New York, NY, USA},
  \bibinfo{pages}{31–44}.
\newblock
\showISBNx{9781450376136}
\urldef\tempurl%
\url{https://doi.org/10.1145/3385412.3386012}
\showDOI{\tempurl}


\bibitem[Nandi et~al\mbox{.}(2021)]%
        {ruler}
\bibfield{author}{\bibinfo{person}{Chandrakana Nandi}, \bibinfo{person}{Max
  Willsey}, \bibinfo{person}{Amy Zhu}, \bibinfo{person}{Yisu~Remy Wang},
  \bibinfo{person}{Brett Saiki}, \bibinfo{person}{Adam Anderson},
  \bibinfo{person}{Adriana Schulz}, \bibinfo{person}{Dan Grossman}, {and}
  \bibinfo{person}{Zachary Tatlock}.} \bibinfo{year}{2021}\natexlab{}.
\newblock \showarticletitle{Rewrite rule inference using equality saturation}.
\newblock \bibinfo{journal}{\emph{Proc. ACM Program. Lang.}}
  \bibinfo{volume}{5}, \bibinfo{number}{OOPSLA}, Article
  \bibinfo{articleno}{119} (\bibinfo{date}{Oct.} \bibinfo{year}{2021}),
  \bibinfo{numpages}{28}~pages.
\newblock
\urldef\tempurl%
\url{https://doi.org/10.1145/3485496}
\showDOI{\tempurl}


\bibitem[Nelson(1979)]%
        {phd79-egraph}
\bibfield{author}{\bibinfo{person}{Charles~Gregory Nelson}.}
  \bibinfo{year}{1979}\natexlab{}.
\newblock \emph{\bibinfo{title}{Techniques for Program Verification}}.
\newblock \bibinfo{thesistype}{Ph.\,D. Dissertation}. \bibinfo{school}{Stanford
  University}.
\newblock


\bibitem[Ng(1993)]%
        {fdlibm}
\bibfield{author}{\bibinfo{person}{Dr. K-C Ng}.}
  \bibinfo{year}{1993}\natexlab{}.
\newblock \bibinfo{title}{{FDLIBM}}.
\newblock
\newblock
\urldef\tempurl%
\url{http://www.netlib.org/fdlibm/readme}
\showURL{%
\tempurl}


\bibitem[Ogita et~al\mbox{.}(2005)]%
        {eft}
\bibfield{author}{\bibinfo{person}{Takeshi Ogita},
  \bibinfo{person}{Siegfried~M. Rump}, {and} \bibinfo{person}{Shin'ichi
  Oishi}.} \bibinfo{year}{2005}\natexlab{}.
\newblock \showarticletitle{Accurate Sum and Dot Product}.
\newblock \bibinfo{journal}{\emph{SIAM Journal on Scientific Computing}}
  \bibinfo{volume}{26}, \bibinfo{number}{6} (\bibinfo{year}{2005}),
  \bibinfo{pages}{1955--1988}.
\newblock
\urldef\tempurl%
\url{https://doi.org/10.1137/030601818}
\showDOI{\tempurl}


\bibitem[{Open Source Geospatial Foundation}({[n.\,d.]})]%
        {proj.org}
\bibfield{author}{\bibinfo{person}{{Open Source Geospatial Foundation}}.}
  \bibinfo{year}{[n.\,d.]}\natexlab{}.
\newblock \bibinfo{title}{{PROJ}}.
\newblock
\newblock
\urldef\tempurl%
\url{https://proj.org/}
\showURL{%
\tempurl}
\newblock
\shownote{Accessed March 2026}.


\bibitem[{OSGeo}({[n.\,d.]})]%
        {osgeo.org}
\bibfield{author}{\bibinfo{person}{{OSGeo}}.}
  \bibinfo{year}{[n.\,d.]}\natexlab{}.
\newblock \bibinfo{title}{Open Source Geospatial Foundation}.
\newblock
\newblock
\urldef\tempurl%
\url{https://www.osgeo.org/}
\showURL{%
\tempurl}
\newblock
\shownote{Accessed March 2026}.


\bibitem[Pal et~al\mbox{.}(2023)]%
        {enumo}
\bibfield{author}{\bibinfo{person}{Anjali Pal}, \bibinfo{person}{Brett Saiki},
  \bibinfo{person}{Ryan Tjoa}, \bibinfo{person}{Cynthia Richey},
  \bibinfo{person}{Amy Zhu}, \bibinfo{person}{Oliver Flatt},
  \bibinfo{person}{Max Willsey}, \bibinfo{person}{Zachary Tatlock}, {and}
  \bibinfo{person}{Chandrakana Nandi}.} \bibinfo{year}{2023}\natexlab{}.
\newblock \showarticletitle{Equality Saturation Theory Exploration \`{a} la
  Carte}.
\newblock \bibinfo{journal}{\emph{Proc. ACM Program. Lang.}}
  \bibinfo{volume}{7}, \bibinfo{number}{OOPSLA2}, Article
  \bibinfo{articleno}{258} (\bibinfo{date}{Oct.} \bibinfo{year}{2023}),
  \bibinfo{numpages}{29}~pages.
\newblock
\urldef\tempurl%
\url{https://doi.org/10.1145/3622834}
\showDOI{\tempurl}


\bibitem[Panchekha et~al\mbox{.}(2015)]%
        {herbie}
\bibfield{author}{\bibinfo{person}{Pavel Panchekha}, \bibinfo{person}{Alex
  Sanchez-Stern}, \bibinfo{person}{James~R. Wilcox}, {and}
  \bibinfo{person}{Zachary Tatlock}.} \bibinfo{year}{2015}\natexlab{}.
\newblock \showarticletitle{Automatically Improving Accuracy for Floating Point
  Expressions} \emph{(\bibinfo{series}{PLDI})}.
\newblock


\bibitem[Park et~al\mbox{.}(2025)]%
        {rlibm-all}
\bibfield{author}{\bibinfo{person}{Sehyeok Park}, \bibinfo{person}{Justin Kim},
  {and} \bibinfo{person}{Santosh Nagarakatte}.}
  \bibinfo{year}{2025}\natexlab{}.
\newblock \showarticletitle{Correctly Rounded Math Libraries without Worrying
  about the Application’s Rounding Mode}.
\newblock \bibinfo{journal}{\emph{Proc. ACM Program. Lang.}}
  \bibinfo{volume}{9}, \bibinfo{number}{PLDI}, Article \bibinfo{articleno}{229}
  (\bibinfo{date}{June} \bibinfo{year}{2025}), \bibinfo{numpages}{24}~pages.
\newblock
\urldef\tempurl%
\url{https://doi.org/10.1145/3729332}
\showDOI{\tempurl}


\bibitem[{PROJ contributors}({[n.\,d.]})]%
        {somerc.cpp}
\bibfield{author}{\bibinfo{person}{{PROJ contributors}}.}
  \bibinfo{year}{[n.\,d.]}\natexlab{}.
\newblock \bibinfo{title}{Swiss Oblique Mercator Projection Implementation}.
\newblock
\newblock
\urldef\tempurl%
\url{https://github.com/OSGeo/PROJ/blob/master/src/projections/somerc.cpp}
\showURL{%
\tempurl}
\newblock
\shownote{Accessed March 2026}.


\bibitem[{PROJ contributors}(2025)]%
        {proj}
\bibfield{author}{\bibinfo{person}{{PROJ contributors}}.}
  \bibinfo{year}{2025}\natexlab{}.
\newblock \bibinfo{title}{{PROJ} Coordinate Transformation Software Library}.
\newblock
\newblock
\urldef\tempurl%
\url{https://doi.org/10.5281/zenodo.5884394}
\showDOI{\tempurl}


\bibitem[Qian et~al\mbox{.}(2026)]%
        {poseidon}
\bibfield{author}{\bibinfo{person}{Siyuan~Brant Qian}, \bibinfo{person}{Vimarsh
  Sathia}, \bibinfo{person}{Ivan~R. Ivanov}, \bibinfo{person}{Jan
  H\"{u}ckelheim}, \bibinfo{person}{Paul Hovland}, {and}
  \bibinfo{person}{William~S. Moses}.} \bibinfo{year}{2026}\natexlab{}.
\newblock \showarticletitle{Thinking Fast and Correct: Automated Rewriting of
  Numerical Code through Compiler Augmentation}. In
  \bibinfo{booktitle}{\emph{Proceedings of the 2026 IEEE/ACM International
  Symposium on Code Generation and Optimization (CGO)}}.
\newblock


\bibitem[Saiki et~al\mbox{.}(2025)]%
        {chassis}
\bibfield{author}{\bibinfo{person}{Brett Saiki}, \bibinfo{person}{Jackson
  Brough}, \bibinfo{person}{Jonas Regehr}, \bibinfo{person}{Jesus Ponce},
  \bibinfo{person}{Varun Pradeep}, \bibinfo{person}{Aditya Akhileshwaran},
  \bibinfo{person}{Zachary Tatlock}, {and} \bibinfo{person}{Pavel Panchekha}.}
  \bibinfo{year}{2025}\natexlab{}.
\newblock \showarticletitle{Target-Aware Implementation of Real Expressions}
  \emph{(\bibinfo{series}{ASPLOS 2025})}. \bibinfo{publisher}{Association for
  Computing Machinery}, \bibinfo{address}{New York, NY, USA},
  \bibinfo{pages}{1069--1083}.
\newblock
\showISBNx{9798400706981}
\urldef\tempurl%
\url{https://doi.org/10.1145/3669940.3707277}
\showDOI{\tempurl}


\bibitem[Saiki et~al\mbox{.}(2021)]%
        {pherbie}
\bibfield{author}{\bibinfo{person}{Brett Saiki}, \bibinfo{person}{Oliver
  Flatt}, \bibinfo{person}{Chandrakana Nandi}, \bibinfo{person}{Pavel
  Panchekha}, {and} \bibinfo{person}{Zachary Tatlock}.}
  \bibinfo{year}{2021}\natexlab{}.
\newblock \showarticletitle{Combining Precision Tuning and Rewriting}. In
  \bibinfo{booktitle}{\emph{2021 IEEE 28th Symposium on Computer Arithmetic
  (ARITH)}}.
\newblock


\bibitem[Sasnauskas et~al\mbox{.}(2018)]%
        {souper}
\bibfield{author}{\bibinfo{person}{Raimondas Sasnauskas}, \bibinfo{person}{Yang
  Chen}, \bibinfo{person}{Peter Collingbourne}, \bibinfo{person}{Jeroen
  Ketema}, \bibinfo{person}{Gratian Lup}, \bibinfo{person}{Jubi Taneja}, {and}
  \bibinfo{person}{John Regehr}.} \bibinfo{year}{2018}\natexlab{}.
\newblock \bibinfo{title}{Souper: A Synthesizing Superoptimizer}.
\newblock
\newblock
\showeprint[arxiv]{1711.04422}~[cs.PL]
\urldef\tempurl%
\url{https://arxiv.org/abs/1711.04422}
\showURL{%
\tempurl}


\bibitem[Schkufza et~al\mbox{.}(2013)]%
        {stoke}
\bibfield{author}{\bibinfo{person}{Eric Schkufza}, \bibinfo{person}{Rahul
  Sharma}, {and} \bibinfo{person}{Alex Aiken}.}
  \bibinfo{year}{2013}\natexlab{}.
\newblock \showarticletitle{Stochastic superoptimization}. In
  \bibinfo{booktitle}{\emph{Proceedings of the Eighteenth International
  Conference on Architectural Support for Programming Languages and Operating
  Systems}} (Houston, Texas, USA) \emph{(\bibinfo{series}{ASPLOS '13})}.
  \bibinfo{publisher}{Association for Computing Machinery},
  \bibinfo{address}{New York, NY, USA}, \bibinfo{pages}{305–316}.
\newblock
\showISBNx{9781450318709}
\urldef\tempurl%
\url{https://doi.org/10.1145/2451116.2451150}
\showDOI{\tempurl}


\bibitem[Schkufza et~al\mbox{.}(2014)]%
        {stoke-float}
\bibfield{author}{\bibinfo{person}{Eric Schkufza}, \bibinfo{person}{Rahul
  Sharma}, {and} \bibinfo{person}{Alex Aiken}.}
  \bibinfo{year}{2014}\natexlab{}.
\newblock \showarticletitle{Stochastic Optimization of Floating Point Programs
  using Tunable Precision} \emph{(\bibinfo{series}{PLDI '14})}.
\newblock


\bibitem[Sibidanov et~al\mbox{.}(2022)]%
        {core-math}
\bibfield{author}{\bibinfo{person}{Alexei Sibidanov}, \bibinfo{person}{Paul
  Zimmermann}, {and} \bibinfo{person}{St\'{e}phane Glondu}.}
  \bibinfo{year}{2022}\natexlab{}.
\newblock \showarticletitle{The {CORE-MATH} Project}. In
  \bibinfo{booktitle}{\emph{2022 IEEE 29th Symposium on Computer Arithmetic
  (ARITH)}}. \bibinfo{pages}{26--34}.
\newblock
\urldef\tempurl%
\url{https://doi.org/10.1109/ARITH54963.2022.00014}
\showDOI{\tempurl}


\bibitem[Singher and Itzhaky(2021)]%
        {thesy}
\bibfield{author}{\bibinfo{person}{Eytan Singher} {and}
  \bibinfo{person}{Shachar Itzhaky}.} \bibinfo{year}{2021}\natexlab{}.
\newblock \showarticletitle{Theory Exploration Powered by Deductive Synthesis}.
  In \bibinfo{booktitle}{\emph{Computer Aided Verification: 33rd International
  Conference, CAV 2021, Virtual Event, July 20–23, 2021, Proceedings, Part
  II}}. \bibinfo{publisher}{Springer-Verlag}, \bibinfo{address}{Berlin,
  Heidelberg}, \bibinfo{pages}{125–148}.
\newblock
\showISBNx{978-3-030-81687-2}
\urldef\tempurl%
\url{https://doi.org/10.1007/978-3-030-81688-9_6}
\showDOI{\tempurl}


\bibitem[Smith et~al\mbox{.}(2024)]%
        {lakeroad}
\bibfield{author}{\bibinfo{person}{Gus~Henry Smith}, \bibinfo{person}{Benjamin
  Kushigian}, \bibinfo{person}{Vishal Canumalla}, \bibinfo{person}{Andrew
  Cheung}, \bibinfo{person}{Steven Lyubomirsky}, \bibinfo{person}{Sorawee
  Porncharoenwase}, \bibinfo{person}{Ren\'{e} Just},
  \bibinfo{person}{Gilbert~Louis Bernstein}, {and} \bibinfo{person}{Zachary
  Tatlock}.} \bibinfo{year}{2024}\natexlab{}.
\newblock \showarticletitle{FPGA Technology Mapping Using Sketch-Guided Program
  Synthesis}. In \bibinfo{booktitle}{\emph{Proceedings of the 29th ACM
  International Conference on Architectural Support for Programming Languages
  and Operating Systems, Volume 2}} (La Jolla, CA, USA)
  \emph{(\bibinfo{series}{ASPLOS '24})}. \bibinfo{publisher}{Association for
  Computing Machinery}, \bibinfo{address}{New York, NY, USA},
  \bibinfo{pages}{416–432}.
\newblock
\showISBNx{9798400703850}
\urldef\tempurl%
\url{https://doi.org/10.1145/3620665.3640387}
\showDOI{\tempurl}


\bibitem[Smith et~al\mbox{.}(2021)]%
        {glenside}
\bibfield{author}{\bibinfo{person}{Gus~Henry Smith}, \bibinfo{person}{Andrew
  Liu}, \bibinfo{person}{Steven Lyubomirsky}, \bibinfo{person}{Scott Davidson},
  \bibinfo{person}{Joseph McMahan}, \bibinfo{person}{Michael Taylor},
  \bibinfo{person}{Luis Ceze}, {and} \bibinfo{person}{Zachary Tatlock}.}
  \bibinfo{year}{2021}\natexlab{}.
\newblock \showarticletitle{Pure tensor program rewriting via access patterns
  (representation pearl)}. In \bibinfo{booktitle}{\emph{Proceedings of the 5th
  ACM SIGPLAN International Symposium on Machine Programming}} (Virtual,
  Canada) \emph{(\bibinfo{series}{MAPS 2021})}. \bibinfo{publisher}{Association
  for Computing Machinery}, \bibinfo{address}{New York, NY, USA},
  \bibinfo{pages}{21–31}.
\newblock
\showISBNx{9781450384674}
\urldef\tempurl%
\url{https://doi.org/10.1145/3460945.3464953}
\showDOI{\tempurl}


\bibitem[Tarjan(1972)]%
        {scc}
\bibfield{author}{\bibinfo{person}{Robert Tarjan}.}
  \bibinfo{year}{1972}\natexlab{}.
\newblock \showarticletitle{Depth-First Search and Linear Graph Algorithms}.
\newblock \bibinfo{journal}{\emph{SIAM J. Comput.}} \bibinfo{volume}{1},
  \bibinfo{number}{2} (\bibinfo{year}{1972}), \bibinfo{pages}{146--160}.
\newblock
\urldef\tempurl%
\url{https://doi.org/10.1137/0201010}
\showDOI{\tempurl}


\bibitem[Tate et~al\mbox{.}(2009)]%
        {eqsat}
\bibfield{author}{\bibinfo{person}{Ross Tate}, \bibinfo{person}{Michael Stepp},
  \bibinfo{person}{Zachary Tatlock}, {and} \bibinfo{person}{Sorin Lerner}.}
  \bibinfo{year}{2009}\natexlab{}.
\newblock \showarticletitle{Equality saturation: a new approach to
  optimization}. In \bibinfo{booktitle}{\emph{Proceedings of the 36th Annual
  ACM SIGPLAN-SIGACT Symposium on Principles of Programming Languages}}
  (Savannah, GA, USA) \emph{(\bibinfo{series}{POPL '09})}.
  \bibinfo{publisher}{Association for Computing Machinery},
  \bibinfo{address}{New York, NY, USA}, \bibinfo{pages}{264–276}.
\newblock
\showISBNx{9781605583792}
\urldef\tempurl%
\url{https://doi.org/10.1145/1480881.1480915}
\showDOI{\tempurl}


\bibitem[Thomas and Bornholt(2024)]%
        {isaria}
\bibfield{author}{\bibinfo{person}{Samuel Thomas} {and} \bibinfo{person}{James
  Bornholt}.} \bibinfo{year}{2024}\natexlab{}.
\newblock \showarticletitle{Automatic Generation of Vectorizing Compilers for
  Customizable Digital Signal Processors}. In
  \bibinfo{booktitle}{\emph{Proceedings of the 29th ACM International
  Conference on Architectural Support for Programming Languages and Operating
  Systems, Volume 1}} (La Jolla, CA, USA) \emph{(\bibinfo{series}{ASPLOS
  '24})}. \bibinfo{publisher}{Association for Computing Machinery},
  \bibinfo{address}{New York, NY, USA}, \bibinfo{pages}{19–34}.
\newblock
\showISBNx{9798400703720}
\urldef\tempurl%
\url{https://doi.org/10.1145/3617232.3624873}
\showDOI{\tempurl}


\bibitem[VanHattum et~al\mbox{.}(2021)]%
        {diospyros}
\bibfield{author}{\bibinfo{person}{Alexa VanHattum}, \bibinfo{person}{Rachit
  Nigam}, \bibinfo{person}{Vincent~T. Lee}, \bibinfo{person}{James Bornholt},
  {and} \bibinfo{person}{Adrian Sampson}.} \bibinfo{year}{2021}\natexlab{}.
\newblock \showarticletitle{Vectorization for digital signal processors via
  equality saturation}. In \bibinfo{booktitle}{\emph{Proceedings of the 26th
  ACM International Conference on Architectural Support for Programming
  Languages and Operating Systems}} (Virtual, USA)
  \emph{(\bibinfo{series}{ASPLOS '21})}. \bibinfo{publisher}{Association for
  Computing Machinery}, \bibinfo{address}{New York, NY, USA},
  \bibinfo{pages}{874–886}.
\newblock
\showISBNx{9781450383172}
\urldef\tempurl%
\url{https://doi.org/10.1145/3445814.3446707}
\showDOI{\tempurl}


\bibitem[Wang et~al\mbox{.}(2020)]%
        {spores}
\bibfield{author}{\bibinfo{person}{Yisu~Remy Wang}, \bibinfo{person}{Shana
  Hutchison}, \bibinfo{person}{Jonathan Leang}, \bibinfo{person}{Bill Howe},
  {and} \bibinfo{person}{Dan Suciu}.} \bibinfo{year}{2020}\natexlab{}.
\newblock \showarticletitle{SPORES: sum-product optimization via relational
  equality saturation for large scale linear algebra}.
\newblock \bibinfo{journal}{\emph{Proc. VLDB Endow.}} \bibinfo{volume}{13},
  \bibinfo{number}{12} (\bibinfo{date}{July} \bibinfo{year}{2020}),
  \bibinfo{pages}{1919–1932}.
\newblock
\showISSN{2150-8097}
\urldef\tempurl%
\url{https://doi.org/10.14778/3407790.3407799}
\showDOI{\tempurl}


\bibitem[{Wikipedia contributors}({[n.\,d.]})]%
        {wiki-invgud}
\bibfield{author}{\bibinfo{person}{{Wikipedia contributors}}.}
  \bibinfo{year}{[n.\,d.]}\natexlab{}.
\newblock \bibinfo{title}{Gudermannian Function}.
\newblock
\newblock
\urldef\tempurl%
\url{https://en.wikipedia.org/wiki/Gudermannian_function}
\showURL{%
\tempurl}
\newblock
\shownote{Accessed March 2026}.


\bibitem[Willsey et~al\mbox{.}(2021)]%
        {egg}
\bibfield{author}{\bibinfo{person}{Max Willsey}, \bibinfo{person}{Chandrakana
  Nandi}, \bibinfo{person}{Yisu~Remy Wang}, \bibinfo{person}{Oliver Flatt},
  \bibinfo{person}{Zachary Tatlock}, {and} \bibinfo{person}{Pavel Panchekha}.}
  \bibinfo{year}{2021}\natexlab{}.
\newblock \showarticletitle{egg: Fast and extensible equality saturation}.
\newblock \bibinfo{journal}{\emph{Proc. ACM Program. Lang.}}
  \bibinfo{volume}{5}, \bibinfo{number}{POPL}, Article \bibinfo{articleno}{23}
  (\bibinfo{date}{Jan.} \bibinfo{year}{2021}), \bibinfo{numpages}{29}~pages.
\newblock
\urldef\tempurl%
\url{https://doi.org/10.1145/3434304}
\showDOI{\tempurl}


\end{thebibliography}

\end{document}